\documentclass[aps,reprint,longbibliography,floatfix,groupedaddress,hidelinks]{revtex4-1}

\usepackage{glossaries} 
\newacronym{rb}{RB}{randomized benchmarking}
\newacronym{gst}{GST}{gate set tomography}
\newacronym{qpt}{QPT}{quantum process tomography}
\newacronym{qec}{QEC}{quantum error correction}
\newacronym{se}{SE}{spin echo}
\newacronym{ptm}{PTM}{Pauli transfer matrix}
\newacronym{me}{ME}{Magnus expansion}
\newacronym{dd}{DD}{dynamical decoupling}
\newacronym{dcg}{DCG}{dynamically corrected gate}
\newacronym{ff}{FF}{filter function}
\newacronym{mc}{MC}{Monte Carlo}
\newacronym{cff}{CFF}{correlation filter function}
\newacronym{qft}{QFT}{quantum Fourier transform}
\newacronym{cp}{CP}{completely positive}
\newacronym{povm}{POVM}{positive operator-valued measure}
\newacronym{iid}{i.i.d.}{independent and identically distributed}

\usepackage{graphicx,epstopdf,xspace,amssymb,amsmath,float}
\usepackage{qcircuit}
\usepackage{listings}
\usepackage{dsfont}
\usepackage[binary-units]{siunitx}
\usepackage[utf8]{inputenc}
\usepackage[english]{babel}
\usepackage{mathtools}
\usepackage{mathrsfs}
\usepackage{physics}
\usepackage[inline]{enumitem}
\usepackage[T1]{fontenc}
\usepackage{csquotes}
\usepackage{xr-hyper}
\PassOptionsToPackage{hyphens}{url}
\usepackage[breaklinks]{hyperref}
\usepackage[nameinlink,capitalise]{cleveref}

\newcommand{\ie}[0]{i.e.\ }
\newcommand{\eg}[0]{e.g.\ }
\newcommand{\viz}[0]{viz.\ }

\newcommand{\cf}[0]{c.f.\ }
\newcommand{\mc}[1]{\ensuremath{\mathcal{#1}}}
\newcommand{\mr}[1]{\ensuremath{\mathrm{#1}}}
\newcommand{\e}[0]{\ensuremath{\mr{e}\xspace}}
\renewcommand{\i}[0]{\ensuremath{\mr{i}\xspace}}
\newcommand{\eps}[0]{\ensuremath{\epsilon}\xspace}
\newcommand{\citer}[1]{Ref.~\citenum{#1}}
\newcommand{\citerr}[2]{Refs.~\citenum{#1} and \citenum{#2}}

\newcommand{\python}{Python\xspace}
\newcommand{\filterfunctions}{\texttt{filter\_functions}\xspace}
\newcommand{\qupulse}{\texttt{qupulse}\xspace}
\newcommand{\qopt}{\texttt{qopt}\xspace}
\newcommand{\numpy}{\texttt{NumPy}\xspace}
\newcommand{\scipy}{\texttt{SciPy}\xspace}
\newcommand{\matplotlib}{\texttt{matplotlib}\xspace}
\newcommand{\qutip}{\texttt{QuTiP}\xspace}
\newcommand{\sparse}{\texttt{sparse}\xspace}
\newcommand{\opteinsum}{\texttt{opt\_einsum}\xspace}
\newcommand{\jupyter}{\texttt{Jupyter}\xspace}
\newcommand{\pulsesequence}{\texttt{PulseSequence}\xspace}
\newcommand{\pulsesequences}{\texttt{PulseSequence}s\xspace}
\newcommand{\qobj}{\texttt{Qobj}\xspace}
\newcommand{\slowprocessor}{AMD FX\texttrademark-6300 processor with six logical cores\xspace}
\newcommand{\fastprocessor}{Intel\textsuperscript{\textregistered}\ Core\texttrademark\ i9-9900K eight-core processor\xspace}

\newcommand{\placeholder}[0]{\:\cdot\:}

\newcommand{\fid}[0]{\ensuremath{\mathcal{F}}\xspace}
\newcommand{\avgfid}[0]{\ensuremath{\fid_\mr{avg}}\xspace}
\newcommand{\entfid}[0]{\ensuremath{\fid_\mr{e}}\xspace}
\newcommand{\infid}[0]{\ensuremath{\mathcal{I}}\xspace}
\newcommand{\leak}[0]{\ensuremath{L}\xspace}
\newcommand{\basis}[0]{\ensuremath{\mc{C}}\xspace}

\newcommand{\entinfid}[0]{\ensuremath{\infid_\mr{e}}\xspace}
\newcommand{\oneoverf}{\ensuremath{\flatfrac{1}{f}}\xspace}

\newcommand{\px}[0]{\ensuremath{\sigma_x}\xspace}
\newcommand{\py}[0]{\ensuremath{\sigma_y}\xspace}
\newcommand{\pz}[0]{\ensuremath{\sigma_z}\xspace}
\newcommand{\eye}[0]{\ensuremath{\mathds{1}}\xspace}
\newcommand{\sts}[0]{\ensuremath{\mathrm{S\mbox{-}T_0}}\xspace}

\newcommand{\ketud}[0]{\mbox{\ensuremath{\ket{\uparrow\downarrow}}}\xspace}
\newcommand{\ketdu}[0]{\mbox{\ensuremath{\ket{\downarrow\uparrow}}}\xspace}

\newcommand{\ketuudd}[0]{\mbox{\ensuremath{\ket{\uparrow\uparrow\downarrow\downarrow}}}\xspace}
\newcommand{\ketdduu}[0]{\mbox{\ensuremath{\ket{\downarrow\downarrow\uparrow\uparrow}}}\xspace}
\newcommand{\ketudud}[0]{\mbox{\ensuremath{\ket{\uparrow\downarrow\uparrow\downarrow}}}\xspace}
\newcommand{\ketdudu}[0]{\mbox{\ensuremath{\ket{\downarrow\uparrow\downarrow\uparrow}}}\xspace}
\newcommand{\ketuddu}[0]{\mbox{\ensuremath{\ket{\uparrow\downarrow\downarrow\uparrow}}}\xspace}
\newcommand{\ketduud}[0]{\mbox{\ensuremath{\ket{\downarrow\uparrow\uparrow\downarrow}}}\xspace}
\newcommand{\brauudd}[0]{\mbox{\ensuremath{\bra{\uparrow\uparrow\downarrow\downarrow}}}\xspace}
\newcommand{\bradduu}[0]{\mbox{\ensuremath{\bra{\downarrow\downarrow\uparrow\uparrow}}}\xspace}

\newcommand{\kpsi}[0]{\mbox{\ensuremath{\ket{\psi}}}\xspace}

\newcommand{\dotHS}[2]{\ensuremath{\expval{#1,#2}}\xspace}
\newcommand{\Hspace}{\ensuremath{\mathscr{H}}\xspace}
\newcommand{\Lspace}{\ensuremath{\mathscr{L}}\xspace}
\newcommand{\ad}{\ensuremath{^\dagger}\xspace}

\newcommand{\gth}[1]{\ensuremath{^{(#1)}}\xspace}

\newcommand{\Li}{\ensuremath{\mathcal{L}}\xspace}
\newcommand{\Lc}{\ensuremath{\mathcal{L}_\mr{c}}\xspace}
\newcommand{\Ln}{\ensuremath{\mathcal{L}_\mr{n}}\xspace}
\newcommand{\Lnt}{\ensuremath{\tilde{\mathcal{L}}_\mr{n}}\xspace}

\newcommand{\Hc}{\ensuremath{H_\mr{c}}\xspace}
\newcommand{\Hn}{\ensuremath{H_\mr{n}}\xspace}
\newcommand{\Hnt}{\ensuremath{\tilde{H}_\mr{n}}\xspace}

\newcommand{\Ue}{\ensuremath{\tilde{U}}\xspace}
\newcommand{\Uc}{\ensuremath{U_\mr{c}}\xspace}

\newcommand{\Rc}{\ensuremath{\bar{\mathcal{B}}}\xspace}
\newcommand{\Ba}{\ensuremath{B_\alpha}\xspace}  
\newcommand{\Bat}{\ensuremath{\tilde{B}_\alpha}\xspace}  
\newcommand{\Bab}{\ensuremath{\bar{B}_\alpha}\xspace}  

\newcommand{\decayamps}[0]{\ensuremath{\Gamma}\xspace}
\newcommand{\freqshifts}[0]{\ensuremath{\Delta}\xspace}
\newcommand{\cumulantfun}[0]{\ensuremath{\mc{K}}\xspace}
\newcommand{\ctrlmat}[0]{\ensuremath{\tilde{\mc{B}}}\xspace}
\newcommand{\liouvP}[0]{\ensuremath{\mc{P}}\xspace}
\newcommand{\liouvQ}[0]{\ensuremath{\mc{Q}}\xspace}
\newcommand{\liouvU}[0]{\ensuremath{\mc{U}}\xspace}
\newcommand{\liouvUc}[0]{\ensuremath{\mc{U}_\mr{c}}\xspace}
\newcommand{\liouvUe}[0]{\ensuremath{\tilde{\mc{U}}}\xspace}

\newcommand{\qp}[0]{\ensuremath{\mc{E}}\xspace}
\newcommand{\dbra}[1]{\mbox{\ensuremath{\left\langle\!\bra{#1}\right.}}\xspace}
\newcommand{\dket}[1]{\mbox{\ensuremath{\left.\ket{#1}\!\right\rangle}}\xspace}
\newcommand{\dbraket}[2]{\mbox{\ensuremath{\left\langle\!\braket{#1}{#2}\!\right\rangle}}\xspace}
\newcommand{\dip}[2]{\dbraket{#1}{#2}}
\newcommand{\ddyad}[2]{\mbox{\ensuremath{\dket{#1}\!\dbra{#2}}}\xspace}

\newcommand{\dop}[2]{\ddyad{#1}{#2}}

\newcommand{\dmatrixel}[3]{\mbox{\ensuremath{\left\langle\!\!\matrixel{#1}{#2}{#3}\!\right\rangle}}\xspace}
\newcommand{\dmel}[3]{\dmatrixel{#1}{#2}{#3}}

\begin{document}

\title{Filter Function Formalism and Software Package to Compute Quantum Processes of Gate Sequences for Classical Non-Markovian Noise}

\author{Tobias Hangleiter}
\email[]{tobias.hangleiter@rwth-aachen.de}
\author{Pascal Cerfontaine}
\email[]{pascal.cerfontaine@rwth-aachen.de}
\author{Hendrik Bluhm}
\affiliation{JARA-FIT Institute for Quantum Information, Forschungszentrum J\"ulich GmbH and RWTH Aachen University, 52074 Aachen, Germany}


\begin{abstract}
Correlated, non-Markovian noise is present in many solid-state systems employed as hosts for quantum information technologies, significantly complicating the realistic theoretical description of these systems. In this regime, the effects of noise on sequences of quantum gates cannot be described by concatenating isolated quantum operations if the environmental correlation times are on the scale of the typical gate durations. The filter function formalism has been successful in characterizing the decay of coherence under the influence of such classical, non-Markovian environments and here we show it can be applied to describe unital evolution within the quantum operations formalism. We find exact results for the quantum process and a simple composition rule for a sequence of operations. This enables the detailed study of effects of noise correlations on algorithms and periodically driven systems. Moreover, we point out the method's suitability for numerical applications and present the open-source \python software package \filterfunctions. Amongst other things, it facilitates computing the noise-averaged transfer matrix representation of a unital quantum operation in the presence of universal classical noise for arbitrary control sequences. We apply the presented methods to selected examples.
\end{abstract}

\maketitle
\section{Introduction}
In the circuit model of quantum computing, computations are driven by applying time-local quantum gates. Any algorithm can be compiled using sequences of one- and two-qubit gates \cite{DiVincenzo1995}. Ideal, error-free gates are represented by unitary transformations, so that simulating the action of an algorithm on an initial state of a quantum computer amounts to simple matrix multiplication. Real implementations are subject to noise that causes decoherence resulting in gate errors. If the noise is uncorrelated between gates, its effect can be described by quantum operations acting as linear maps on density matrices, even when several gates are concatenated.
A closely related approach is the use of a Master equation in Lindblad form \cite{Lindblad1976}, which governs the dynamics of density matrices under the influence of Markovian noise with a flat power spectral density.

Yet many physical systems used as hosts for qubits do not satisfy the condition of uncorrelated noise. One example frequently encountered in solid state systems is that of \oneoverf noise, which in principle contains arbitrarily long correlation times. It emerges for instance as flux noise in superconducting qubits and electrical noise in quantum dot qubits \cite{Brownnutt2015,Kumar2016,Yoneda2018,Paladino2014}. Whereas simple approaches exist to treat for example quasistatic noise, which corresponds to perfectly correlated noise (\ie a spectrum with weight only at zero frequency), they cannot be applied to \oneoverf noise because of the wide distribution of correlation times it contains. Thus, there is a gap in the mathematical descriptions of gate operations for noises with arbitrary power spectra that exist between the extremal cases of perfectly flat (white) and sharply peaked (quasistatic) spectra. To capture experimentally relevant effects important to understand the capabilities of quantum computing systems, a universally applicable formalism is hence desirable. For example, one may expect the fidelity requirements for quantum error correction to be more stringent for correlated noise as errors of different gates can interfere constructively \cite{Ng2009}. On the other hand, it might also be possible to use correlation effects to one's benefit, attenuating decoherence by cleverly constructing the gate sequences in algorithms.

As experimental platforms begin to approach fidelity limits set by employing primitive pulse schemes \cite{Veldhorst2014,Debnath2016,Yoneda2018} and detailed knowledge about noise sources and spectra in solid-state systems becomes available \cite{Dial2013,Quintana2017,Malinowski2017spectroscopy}, control pulse optimizations tailored towards specific systems will be required to further push fidelities beyond the error correction threshold \cite{Barends2014,Blume-kohout2017}. This calls for flexible and generically applicable tools as a basis for the numerical optimization of pulses as well as the detailed analysis of the quantum processes they effect. In order to obtain a useful description also for gate operations that decouple from leading orders of noise, such as \glspl{dcg} \cite{Khodjasteh2009}, beyond leading order or exact results are required.

In an accompanying publication \cite{Cerfontaine2021} we presented a formalism based on filter functions and the \gls{me} that addresses these needs and limitations of the canonical master equation approach for correlated noise. Specifically, we showed how process descriptions can be obtained perturbatively for arbitrary classical noise spectra and derived a concatenation rule to obtain the filter function of a sequence of gates from those of the individual gates. This paper extends these results.

\Glspl{ff} were originally introduced to describe the decay of phase coherence under \gls{dd} sequences \cite{Kofman2001,Martinis2003,Uhrig2007,Cywinski2008} consisting of wait times and perfect $\pi$-pulses. The formalism facilitated recognizing these sequences as band-pass filters that allow for probing the environmental noise characteristics of a quantum system through noise spectroscopy \cite{Alvarez2011,Bylander2011,Paz-Silva2017,Malinowski2017spectroscopy} or optimizing sequences to suppress specific noise bands \cite{Biercuk2009,Uys2009,Soare2014,Malinowski2017}. It can also be extended to fidelities of gate operations for single \cite{Green2012,Green2013} or multiple \cite{Gungordu2018,Ball2020} qubits using the \gls{me} \cite{Magnus1954,Blanes2009} as well as more general \gls{dd} protocols \cite{Paz-Silva2014}. The works by \citet{Green2013} and \citet{Clausen2010} also introduced the notion of the control matrix as a quantity closely related to the canonical filter function that is convenient for calculations. In this context, the formalism's capability to predict fidelities of gate implementations has been identified and experimentally tested \cite{Green2013,Kabytayev2014,Soare2014,Ball2016}. Recently, it has also proved useful in assessing the performance requirements for classical control electronics \cite{VanDijk2018}.

While analytic approaches allow for the calculation of filter functions of arbitrary quantum control protocols in principle, it is in practice often a tedious task to determine analytic solutions to the integrals involved if the complexity of the applied wave forms goes beyond simple square pulses or extend to multiple qubits. Moreover, one does not always have a closed-form expression of the control at hand, such as is the case for numerically optimized control pulses. This calls for a numerical approach which, while giving up some of the insights an analytical form offers, is universally applicable and eliminates the need for laborious analytic calculations.

Here, we build and extend upon our accompanying work of \citer{Cerfontaine2021} and that of \citer{Green2013} to show that the formalism can be recast within the framework of stochastic Liouville equations by means of the cumulant expansion \cite{Kubo1962,Kubo1963} which, for Gaussian noise, entails exact results for the quantum process of an arbitrary control operation using only first and second order terms of the \gls{me} \cite{Magnus1954}. Moreover, due to the fact that the \gls{me} retains the algebraic structure of the expanded quantity \cite{Blanes2009} we are able to separate decoherent and coherent contributions to the process. We give explicit methods to evaluate these terms for piecewise-constant control pulses. Moreover, we show that the formalism naturally lends itself as a tool for numerical calculations and present the \filterfunctions \python software package that enables calculating the filter function of arbitrary, piecewise constant defined pulses \cite{software}. On top of providing methods to handle individual quantum gates, the package also implements the concatenation operation as well as parallelized execution of pulses on different groups of qubits, allowing for a highly modular and hence computationally powerful treatment of quantum algorithms in the presence of correlated noise. Given an arbitrary, classical noise spectral density, it can be used to calculate a matrix representation of the error process. From this matrix one can extract average gate fidelities, transition probabilities, and leakage rates as we derive below. To simplify adaptation the software's API is strongly inspired by and compatible with \qutip \cite{Johansson2013} as well as \qopt \cite{Teske2021}. This allows users to use these packages in conjunction. Assessing the computational performance, we show that our method outperforms Monte Carlo simulations for single gates. New analytic results applicable to periodic Hamiltonians and employing the concatenation property make this advantage even more pronounced for sequences of gates. To highlight the main software features, we show example applications below.

We provide this package in the expectation that it will be a useful tool for the community. Besides recasting and expanding on our earlier introduction of the formalism in \citer{Cerfontaine2021}, the present paper is intended to provide an overview of the software and its capabilities. It is structured as follows: In \cref{sec:theory} we derive an exact expression for unital quantum operations in the presence of non-Markovian Gaussian noise and lay out how it may be evaluated using the filter function formalism. We review the concatenation of quantum operations shown in \citer{Cerfontaine2021} and furthermore adapt the method by \citet{Green2013} to calculate the filter function of an arbitrary control sequence numerically. We will specifically focus on computational aspects of the formalism and lay out how to compute various quantities of interest. Moreover, we classify its computational complexity for calculating average gate fidelities and remark on simplifications that allow for drastic improvements in performance in certain applications. In \cref{sec:software}, we introduce the software package by outlining the programmatic structure and giving a brief overview over the API. Lastly, in \cref{sec:examples}, we show the application of the software by means of four examples that highlight various features of the formalism and its implementation. Therein, we first demonstrate that the formalism can predict average gate fidelities for complex two-qubit quantum gates in agreement with computationally much more costly Monte Carlo calculations. Next, we show how it can be applied to periodically driven systems to efficiently analyze Rabi oscillations. We finally establish the formalism's ability to predict deviations from the simple concatenation of unitary gates for sequences and algorithms in the presence of correlated noise by simulating a randomized benchmarking experiment as well as assembling a \gls{qft} circuit from numerically optimized gates. We conclude by briefly remarking on possible future application and extension of our method in \cref{sec:considerata}.

Throughout the paper we will denote operators by Roman font, \eg $U$, and quantum operations and their representations as transfer matrices by calligraphic font, \eg \liouvU, which we also use for the control matrix \ctrlmat to emphasize its innate connection to a transfer matrix. For consistency, a unitary quantum operation will share the same character as the corresponding unitary operator. An operator in the interaction picture will furthermore be designated by an overset tilde, \eg $\tilde{H} = U\ad H U$ with $U$ the unitary operator defining the co-moving frame. Definitions of new quantities on the left and right side of an equality are denoted by $\coloneqq$ and $\eqqcolon$, respectively. We use a central dot ($\placeholder$) as a placeholder in some definitions of abstract operators such as the Liouvillian, denoted by $\mc{L}\coloneqq\comm{H}{\placeholder}$, which is to be understood as the commutator of the corresponding Hamiltonian $H$ and the operator that $\mc{L}$ acts on. The identity matrix is denoted by \eye and its dimension always inferred from context. Furthermore, we will use Greek letters for indices that correspond to noise operators in order to distinguish them clearly from those that correspond to basis or matrix elements. Lastly, we work in units where $\hbar =  1$.

\section{Filter function formalism for unital quantum operations}\label{sec:theory}
We begin the theoretical part of this article by showing how a superoperator matrix representation of the error process, the \enquote{error transfer matrix}, of a unital quantum operation can be computed from the control matrix of the pulse implementing the operation. The control matrix relates the operators through which noise couples into the system to a set of basis operators in the interaction picture and we detail how it can be calculated in a relatively efficient manner for two different situations. First, we consider a sequence of gates whose control matrices have been precomputed. Second, we lay out how the control matrix can be obtained from scratch under the assumption of piecewise constant control, which is often convenient for approximating continuous pulse shapes. Other wave forms can be dealt with analogously by solving the corresponding integrals. We then move on to show how several quantities of interest can be extracted and present optimized strategies for computing the central objects of the formalism.

\subsection{Transfer matrix representation of quantum operations}\label{sec:theory:transfer_matrix}
\subsubsection{Brief review of quantum operations and superoperators}
The quantum operations formalism provides a general framework for the description of open quantum systems \cite{Kraus1983,NielsenChuang2011}. It forms the mathematical basis for \gls{qpt} \cite{Chuang1997,Poyatos1997} as well as \gls{gst} \cite{Blume-Kohout2013,Greenbaum2015} and has also been extensively employed in the context of \gls{rb} \cite{Magesan2012,Kimmel2014}. Several different representations of quantum operations exist. While all of them are equivalent one typically chooses the most convenient for the problem at hand. For an overview of the most commonly used representations see \citer{Greenbaum2015} and for matrix representations in particular \citer{Nambu2008} and the references therein. In this work we employ the Liouville representation, to the best of our knowledge first formalized by \citet{Fano1957}, to profit from its simple properties under composition. It is also known as the transfer matrix representation and we will use the terms interchangeably below. We now briefly review the concept and refer the reader to the literature for further details. Concretely, the Liouville representation of an operation $\qp: \rho\rightarrow\qp(\rho)$ acting on density operators in a Hilbert space \Hspace of dimension $d$ is given by
\begin{equation}\label{eq:liouville_representation}
    \qp_{ij}\doteq\tr(C_i\ad\qp(C_j))
\end{equation}
with an operator basis $\basis=\lbrace C_0, C_1, \dotsc, C_{d^2-1}\rbrace$ for the space of linear operators over \Hspace, $\mathsf{L}(\Hspace)$, orthonormal with respect to the Hilbert-Schmidt product $\dotHS{A}{B}\coloneqq\tr(A\ad B)$. In the case that the operator basis corresponds to the Pauli matrices \cref{eq:liouville_representation} is known as the Pauli transfer matrix (PTM). The operation \qp is thus associated with a $d^2\times d^2$ matrix in \emph{Liouville space} \Lspace that describes its action as the degree to which the $j$-th basis element is mapped onto the $i$-th. On \Lspace one can identify a set of basis kets $\lbrace\dket{C_i}\rbrace_{i=0}^{d^2-1} = \lbrace\dket{i}\rbrace_{i=0}^{d^2-1}$ isomorphic to the operators $C_i$ (and correspondingly bras $\dbra{i}$ to the adjoint $C_i\ad$) as well as the inner product $\dip{i}{j} = \dotHS{C_i}{C_j}$. As the vectors $\dket{i}$ form an orthonormal basis, any operator on \Hspace can be written as a vector on \Lspace, $\dket{A} = \sum_i\dket{i}\dip{i}{A}$, whereas a superoperator on \Hspace becomes a matrix on \Lspace, see \cref{eq:liouville_representation}. It can then be shown that density operators represented by vectors are propagated by transfer matrices so that the action of a quantum operation \qp on a density operator $\rho$ is given by $\dket{\qp(\rho)} = \qp\dket{\rho} = \sum_{ij}\dket{i}\dmel{i}{\qp}{j}\dip{j}{\rho}$. Thus, the composition of two operations $\qp_1$ and $\qp_2$ corresponds to matrix multiplication in Liouville space, $[\qp_2\circ\qp_1]_{ik} = \sum_j[\qp_2]_{ij}[\qp_1]_{jk}$, a property which makes the representation particularly attractive for sequences of operations. Although from a numerical perspective the computational complexity scales unfavorably with the system dimension $d$ (\cf \cref{sec:performance:complexity}),  we will employ the Liouville representation for its transparent interpretation and concise behavior under composition in the following analytical considerations. Lastly, we note that for $C_0\propto\eye$, trace-preservation and unitality are encoded in the relations $\qp_{0j} = \delta_{0j}$ and $\qp_{j0} = \delta_{j0}$, respectively.

\subsubsection{Liouville representation of the error channel}\label{sec:theory:transfer_matrix:derivation}
We will now derive an expression for the quantum process of a quantum gate in the presence of arbitrary classical noise. As a single realization of a classical noise generates strictly unitary dynamics, we will be interested in the expectation value of the dynamics over many such realizations, which will lead to a quantum process including decoherence. If the noise is additionally Gaussian, these results are exact and therefore apply without restrictions to arbitrarily large noise strength as well as to gates that partially decouple from noise. For such \glspl{dcg} or \gls{dd} sequences \cite{Khodjasteh2009,Cywinski2008} higher order terms can become dominant. In the case that the environment is not strictly Gaussian, our approach becomes perturbative and we recover the results presented in \citer{Cerfontaine2021}. As most of our discussion later on in this article will focus on the leading order approximation, readers not interested in the full generality may refer to that publication for a less general but perhaps more accessible derivation and skip ahead to \cref{sec:theory:decay_amplitudes}.

The difference is that in \citer{Cerfontaine2021}, the Magnus expansion is applied to the solution of the Schr\"odinger equation, whereas the approach presented here is based on the theory of stochastic Liouville equations and the cumulant expansion \cite{Kubo1962,Kubo1963}. In the filter function context, the cumulant expansion has been used to express the decay of the off-diagonal terms of a single-qubit density matrix in \citer{Cywinski2008}. More recently, \citet{Paz-Silva2014} employed it in conjunction with the \gls{me} to obtain the matrix elements of the perturbed density operator after a time $T$ of noisy evolution. In \citer{Yang2019}, the authors made use of the cumulant expansion and stochastic Liouville equations for the purpose of gate optimization. Here, we combine different aspects of these works and make the connection to the quantum operations formalism by determining the noise-averaged error propagator in the Liouville representation. This form completely characterizes the error process and hence allows for detailed insight into the decoherence mechanisms of the operation.

Concretely, we consider a system described by the stochastic Hamiltonian
\begin{gather}
    H(t) = \Hc(t) + \Hn(t),\label{eq:hamiltonian} \\
    \Hn(t) = \sum_\alpha b_\alpha(t) B_\alpha(t). \label{eq:hamiltonian:noise}
\end{gather}
$\Hc(t)$ is implemented by the experimentalist to generate the desired control operation during the time $t\in [0, \tau]$ and $\Hn(t)$ describes classical fluctuating noise environments $b_\alpha(t)\in\mathbb{R}$ that couple to the quantum system via the Hermitian noise operators $B_\alpha(t)\in\mathsf{L}(\Hspace)$. These may carry a general, deterministic time dependence and without loss of generality, we can require them to be traceless since any contributions proportional to the identity do not contribute to noisy evolution in any case \footnote{The identity commutes with the control Hamiltonian at all times and hence does not generate any evolution in the interaction picture in which we work later on (\cf \cref{eq:cumulant:truncated:substituted})}. The $b_\alpha(t)$ are random variables drawn from (not necessarily Gaussian) distributions with zero mean that are assumed to be \gls{iid} both with respect to repetitions of the experiment. Note that this concept of independence does not preclude correlations between different noise sources $\alpha\neq\beta$ nor between one noise source at different times $t\neq t'$, but only serves to obtain a well-defined ensemble average. Lastly, to be able to later on relate the correlation functions of the $b_\alpha(t)$ to their spectral density, we require the noise fields to be wide-sense stationary, meaning that their correlation function depends only on the time difference.

For noise operators without explicit time dependence, \cref{eq:hamiltonian:noise} constitutes a universal decomposition as can be seen by choosing the $B_\alpha$ from an orthonormal basis for $\mathsf{L}(\Hspace)$. To motivate the time-dependent form of \cref{eq:hamiltonian:noise}, assume the true Hamiltonian is a function of a set of noisy parameters $\vec{\tilde{\lambda}}(t) = \vec{\lambda}(t) + \vec{\delta\lambda}(t)$ where $\vec{\delta\lambda}(t) = \text{vec}(\{b_\alpha(t)\}_\alpha)$ are the stochastic variables. Expanding the Hamiltonian in an orthonormal operator basis yields $H(\vec{\tilde{\lambda}}(t)) = \sum_\alpha f_\alpha(\vec{\lambda}(t), \vec{\delta\lambda}(t)) B_\alpha$. In general, however, the expansion coefficients $f_\alpha$ will be arbitrary functions of both the deterministic parameters $\vec{\lambda}(t)$ and the stochastic noises $\vec{\delta\lambda}(t)$, which prohibits a factorized form like \cref{eq:hamiltonian:noise}. We can address this problem by first expanding $H$ around $\vec{\lambda}(t)$ for small fluctuations $\vec{\delta\lambda}(t)$. Then, the Hamiltonian approximately becomes $H(\vec{\tilde{\lambda}}(t)) \approx H(\vec{\lambda}(t)) + \vec{\delta\lambda}(t)\cdot\vec{\nabla}_{\lambda} H(\vec{\lambda}(t))$, where we can define the control Hamiltonian as $\Hc(t)\coloneqq H(\vec{\lambda}(t))$. Expanding the second term in the operator basis now results in the form \eqref{eq:hamiltonian:noise} for the noise Hamiltonian as it is linear in $\vec{\delta\lambda}(t)$ and the deterministic time dependence is contained in $\vec{\nabla}_{\lambda} H(\vec{\lambda}(t))$ alone.

This permits us to model complex relations between physical noise sources and the noise operators that capture the coupling to the quantum system, arising for example through control hardware or effective Hamiltonians obtained from \eg Schrieffer-Wolff transformations. While the linearization is in most cases an approximation, it does not impose significant constraints since the noise is typically weak compared to the control \footnote{The same argument forms the basis for the perturbative approach for non-Gaussian noise.}. As an example, we could capture a dependence of the device sensitivity on external controls (see also \citer{Gungordu2018}). In a widely used setting electrons confined in solid-state quantum dots are manipulated using the exchange interaction $J$ that depends non-linearly on the potential difference $\eps$ between two dots. Since the dominant physical noise source affecting this control is charge noise, one could include the effect on $J(\eps)$ to first order with $s_\eps(t)=\pdv*{J(\eps(t))}{\eps(t)}$ so that $\Hn(t) = b_\eps(t) B_\eps(t)  =  b_\eps(t) s_\eps(t) B_\eps$ for some operator $B_\eps$ which represents the exchange coupling.

We proceed in our derivation by noting that the control Hamiltonian $\Hc$ gives rise to the noise-free Liouville--von Neumann equation 
\begin{equation}\label{liouville-von-neumann}
    \dv{\rho(t)}{t} = -\i\comm{\Hc(t)}{\rho(t)} =  -\i\Lc(t)\rho(t)
\end{equation}
on the Hilbert space \Hspace with the Liouvillian superoperator $\Lc(t)$ representing the control. Analogous to the Schr\"odinger equation we may also write this differential equation in terms of time evolution superoperators (superpropagators), $\dv*{\liouvUc(t)}{t} = -\i\Lc(t)\liouvUc(t)$ where the action of \liouvUc on a state $\rho$ is to be understood as $\liouvUc\!: \rho\rightarrow\Uc\rho\Uc\ad$ with \Uc the usual time evolution operator satisfying the corresponding Schr\"odinger equation. This allows us to write the superpropagator for the total Liouvillian $\Li = \Lc + \Ln$ as $\liouvU(t) = \liouvUc(t)\liouvUe(t)$ where the unitary error superpropagator $\liouvUe(t)$ contains the effect of a specific noise realization in \cref{eq:hamiltonian:noise}. Next, we transform the noise Liouvillian \Ln to the interaction picture with respect to the control Liouvillian $\Lc$ so that $\liouvUe(t)$ satisfies the modified Liouville equation
\begin{gather}
    \dv{\liouvUe(t)}{t} = -\i\Lnt(t)\liouvUe(t),    \label{eq:le:interaction_picture} \\
    \Lnt(t) = \liouvUc^\dagger(t)\Ln(t)\liouvUc(t). \label{eq:liouvillian:interaction_picture}
\end{gather}
\Cref{eq:le:interaction_picture} may be formally solved using the \acrlong{me} \cite{Magnus1954} so that at time $t=\tau$
\begin{equation}\label{eq:error_propagator}
    \liouvUe(\tau) = \exp(-\i\tau\Li_\mr{eff}(\tau))
\end{equation}
with $\Li_\mr{eff}(\tau) = \sum_{n=1}^\infty\Li_{\mr{eff},n}(\tau)$. A sufficient criterion for the convergence of the expansion is given by \citet{Moan1999} as $\int_0^\tau\dd{t}\norm*{\Lnt(t)} < \pi$ where $\norm{\placeholder} = \sqrt{\dotHS{\placeholder}{\placeholder}}$ is the Frobenius (Hilbert-Schmidt) norm. The first and second terms of the \gls{me} are given by \cite{Magnus1954,Blanes2009}
\begin{subequations}\label{eq:magnus_expansion}
\begin{align}
    \Li_\mr{eff,1}(\tau) &= \frac{1}{\tau}\int_0^\tau\dd{t}\Lnt(t), \label{eq:magnus_expansion:1}\\
    \Li_\mr{eff,2}(\tau) &= -\frac{\i}{2\tau}\int_0^\tau\dd{t_1}\int_0^{t_1}\dd{t_2}\comm{\Lnt(t_1)}{\Lnt(t_2)}. \label{eq:magnus_expansion:2}
\end{align}
\end{subequations}
The $n$-th term of the expansion contains $n$ factors of the noise variables $b_\alpha(t)$ and scales with $n$ factors of the control duration $\tau$, suggesting that higher-order terms can be neglected if their product is small. In the Bloch sphere picture this corresponds to requiring that the angle by which the Bloch vector is rotated away from its intended trajectory due to the noise be small. Below, we will use the parameter $\xi$ to denote the magnitude of this deviation. It is properly defined in \cref{appsec:convergence:magnus_expansion} where also bounds for the convergence of the \gls{me} are discussed. Here, we only state that $\Li_{\mr{eff},n}\sim\xi^n$ (see also \citer{Green2013}).

We have suggestively written the \gls{me} in terms of an effective Liouvillian $\Li_\mr{eff} = \comm{H_\mr{eff}}{\placeholder}$ to interpret it as the generator of a \emph{time}-averaged evolution of a single noise realization up to time $\tau$. In order to obtain the \emph{ensemble}-averaged evolution of many realizations of the stochastic Hamiltonian in \cref{eq:hamiltonian:noise}, we apply the cumulant expansion to \liouvUe (see also \citerr{Beaudoin2015}{Willick2018}),
\begin{equation}\label{eq:cumulant_expansion}
    \ev*{\liouvUe(\tau)} = \ev{\exp(-\i\tau\Li_\mr{eff}(\tau))} \eqqcolon \exp\cumulantfun(\tau)
\end{equation}
with $\ev{\placeholder}$ denoting the ensemble average \footnote{
The ensemble average represents the expectation value over identical repetitions of an operation in an experiment. It can be taken to be a spatial ensemble of many identical systems, \eg an NMR system, or, for ergodic systems, a time ensemble of a single system under stationary noise as would be the case for a single spin measured repeatedly, for instance.
} and the cumulant function \cite{Kubo1962}
\begin{align}
    \cumulantfun(\tau) &= \sum_{k=1}^\infty\frac{(-\i\tau)^k}{k!}\ev{\Li_\mr{eff}(\tau)^k}_\mr{c} \\
                       &= \sum_{k=1}^\infty\frac{(-\i\tau)^k}{k!}\ev{\left[\sum_{n=1}^\infty \Li_{\mr{eff},n}(\tau)\right]^k}_\mr{c}. \label{eq:cumulant}
\end{align}
The notation $\ev{\placeholder}_\mr{c}$ denotes the cumulant average which prescribes a certain averaging operation. The first cumulant of a set of random variables $\{X_i(t)\}_i$ is simply the expectation value, $\ev{X_i(t)}_\mr{c} = \ev{X_i(t)}$, whereas the second cumulant corresponds to the covariance, $\ev{X_i(t)X_j(t)}_\mr{c} = \ev{X_i(t)X_j(t)} - \ev{X_i(t)}\ev{X_j(t)}$. Remarkably, third and higher-order cumulants vanish for Gaussian processes \cite{Kubo1963,Szankowski2017}, making \cref{eq:cumulant} exact by truncating the sums already at $k = 2$ and $n = 2$. In this case, the convergence radius of the \gls{me} becomes infinite. The terms with $k = n = 2$ do not contribute as they involve fourth-order cumulants. Since furthermore we assume that the noise fields $b_\alpha(t)$ have zero mean, also the terms with $k = n = 1$ vanish and $\ev{X_i(t)X_j(t)}_\mr{c}  =  \ev{X_i(t)X_j(t)}$. We can hence write the cumulant function succinctly as
\begin{equation}
    \cumulantfun(\tau) = - \i\tau\ev{\Li_{\mr{eff},2}(\tau)} - \frac{\tau^2}{2}\ev{\Li_{\mr{eff},1}(\tau)^2}  \label{eq:cumulant:truncated}.
\end{equation}
\Cref{eq:cumulant_expansion,eq:cumulant:truncated} allow us to exactly compute the full quantum process $\ev*{\liouvUe}\!: \rho\rightarrow\ev*{\liouvUe(\rho)}$ for Gaussian noise with arbitrary spectral density and power. For non-Gaussian noise these expressions are approximate up to $\order{\xi^2}$ and higher order terms include both higher orders of the \gls{me} and the cumulant expansion. Inspecting \cref{eq:cumulant:truncated}, we observe that the first term is anti-Hermitian as it is a pure Magnus term (remember that the \gls{me} preserves algebraic structure to every order) and thus generates unitary, coherent time evolution. Conversely, the second term is Hermitian and thus generates decoherence \footnote{In the Liouville representation, the first term is an antisymmetric matrix that generates a rotation and the second a symmetric matrix that generates a deformation of the generalized, $d^2-1$-dimensional Bloch sphere.}. The former is more difficult to compute than the latter because the second order of the \gls{me}, \cref{eq:magnus_expansion:2}, contains nested time integrals. Arguments can be made \cite{Cerfontaine2021}, however, that for single gates in an experimental context the coherent errors captured by this term can be calibrated out to a large degree \cite{Cerfontaine2019gsc,Kimmel2015}. Moreover, many of the central quantities of interest that can be extracted from the quantum process, among which are gate fidelities and certain measurement probabilities, are functions of only the diagonal elements of \cumulantfun. By virtue of the antisymmetry of the second order terms, they do not contribute to these quantities to leading order as we show in \cref{sec:theory:derived_quantities}.

While we will also lay out how to compute the second order, our discussion will therefore focus on contributions from the incoherent term below. As it turns out, this term can be computed using a filter function formalism based on that by \citet{Green2013}. To see this, we insert the explicit forms of the \gls{me} given in \cref{eq:magnus_expansion} and the noise Hamiltonian given in \cref{eq:hamiltonian:noise} into \cref{eq:cumulant:truncated}. Together with $\comm{\mc{L}}{\mc{L}'} = \comm{\comm{H}{H'}}{\placeholder}$ and $\mc{L}\mc{L}' = \comm{H}{\comm{H'}{\placeholder}}$, we find that
\begin{widetext}
\begin{align}
    \cumulantfun(\tau) = -\frac{1}{2}\sum_{\alpha\beta}\Biggl(\int_0^\tau\dd{t_1}\int_0^{t_1}\dd{t_2}
                              \expval{b_\alpha(t_1) b_\beta(t_2)} & \comm{\comm{\tilde{B}_\alpha(t_1)}{\tilde{B}_\beta(t_2)}}{\placeholder} \notag\\
                                                             +\int_0^\tau\dd{t_1}\int_0^\tau\dd{t_2}
                              \expval{b_\alpha(t_1) b_\beta(t_2)} & \comm{\tilde{B}_\alpha(t_1)}{\comm{\tilde{B}_\beta(t_2)}{\placeholder}}\Biggr), \label{eq:cumulant:truncated:substituted}
\end{align}
\end{widetext}
where $\Bat(t) = \Uc\ad(t)\Ba(t)\Uc(t)$ are the noise operators of \cref{eq:hamiltonian:noise} in the interaction picture. $\ev{b_\alpha(t_1)b_\beta(t_2)}$ is the cross-correlation function of noise sources $\alpha$ and $\beta$ which we will later relate to the spectral density. For now, we stay in the time domain and introduce an orthonormal and Hermitian operator basis for the Hilbert space \Hspace to define the Liouville representation,
\begin{equation}\label{eq:basis}
    \basis = \lbrace C_k\in\mathsf{L}(\Hspace): C_k\ad = C_k\:\text{and}\:\tr(C_k C_l) = \delta_{kl}\rbrace_{k=0}^{d^2-1},
\end{equation}
where we choose $C_0 = \flatfrac{\eye}{d^{\flatfrac{1}{2}}}$ for convenience so that the remaining elements are traceless. In order to separate the commutators from the time-dependence and hence the integral in \cref{eq:cumulant:truncated:substituted}, we expand the noise operators in this basis so that
\begin{equation}\label{eq:noise_operators:expanded}
    \Bat(t) \eqqcolon \sum_k\ctrlmat_{\alpha k}(t) C_k.
\end{equation}
The expansion coefficients $\ctrlmat_{\alpha k}(t)\in\mathbb{R}$ are given by the inner product of a noise operator in the interaction picture on the one hand and a basis element on the other:
\begin{equation}\label{eq:control_matrix}
    \ctrlmat_{\alpha k}(t) = \langle\Bat(t), C_k\rangle  = \tr(\Uc\ad(t)\Ba(t)\Uc(t)C_k).
\end{equation}
In line with \citet{Green2013}, we call these coefficients the control matrix (see also \citerr{Byrd2002}{Clausen2010}). In the transfer matrix (superoperator) picture we can take up the following interpretation for the control matrix by virtue of the cyclicity of the trace: it describes a mapping of a state, represented by the basis element $C_k$ and subject to the control operation $\liouvUc(t): C_k\rightarrow\Uc(t) C_k\Uc\ad(t)$, onto the noise operator $\Ba(t)$. That is, we can write the $\alpha$-th row of the control matrix as $\langle\!\langle{\Bat(t)}\rvert = \dbra{\Ba(t)}\liouvUc(t)$. In this connection lies the power of the \gls{ff} formalism as will become clear shortly; we can first determine the ideal evolution without noise and subsequently evaluate the error process by linking the unitary control operation to the noise operators.

Having expanded the noise operators in the basis \basis, we can already anticipate that upon substituting them, \cref{eq:cumulant:truncated:substituted} will separate into a time-dependent part involving on one hand the control matrix and cross-correlation functions and on the other a time-independent part involving commutators of basis elements. This will simplify our calculations in the following. To see this, we recall the definition of the Liouville representation in \cref{eq:liouville_representation} and apply it to the cumulant function so that $\cumulantfun_{ij} = \tr(C_i\cumulantfun[C_j])$, where the notation $\cumulantfun[C_j]$ means substituting $C_j$ for the placeholder $\placeholder$ in the commutators in \cref{eq:cumulant:truncated:substituted} and we suppressed the time argument for legibility. Finally, we insert the expanded noise operators given by \cref{eq:noise_operators:expanded} and obtain the Liouville representation of the cumulant function,
\begin{equation}
    \cumulantfun_{ij}(\tau) \eqqcolon -\frac{1}{2}\sum_{\alpha\beta} \sum_{kl}\left(
        f_{ijkl}\freqshifts_{\alpha\beta,kl} + g_{ijkl}\decayamps_{\alpha\beta,kl}
    \right). \label{eq:cumulant:truncated:liouville}
\end{equation}
Here, we captured the ordering of the noise operators due to the commutators in \cref{eq:cumulant:truncated:substituted} in the coefficients $f_{ijkl}$ and $g_{ijkl}$. These are trivial functions of the fourth order trace tensor
\footnote{Note the similarity to the relationship of a transfer matrix with the $\chi$--matrix, $\qp_{ij} = \sum_{kl}\chi_{kl} T_{i k j l}$, with $\chi_{kl}$ defined by $\qp(\rho) = \sum_{kl}\chi_{kl} C_k\rho C_l$ or, in terms of the Kraus operators $K_i$ of the quantum operation, $\chi_{kl} = \sum_i \tr(K_i C_k) \mr{tr}(K_i\ad C_l)  = \left[\sum_i\dop{K_i}{K_i}\right]_{kl}$ \cite{Greenbaum2015}}
\begin{equation}\label{eq:trace_tensor}
    T_{ijkl} = \tr(C_i C_j C_k C_l)
\end{equation}
given by
\begin{subequations}
\begin{align}\label{eq:structure_constants}
    f_{ijkl} &= T_{klji} - T_{lkji} - T_{klij} + T_{lkij}\qand \\
    g_{ijkl} &= T_{klji} - T_{kjli} - T_{kilj} + T_{kijl}.
\end{align}
\end{subequations}
Furthermore, we introduced the frequency (Lamb) shifts \freqshifts and decay amplitudes \decayamps which contain all information on the noise and qubit dynamics as captured by the control matrix $\ctrlmat(t)$:
\begin{align}
    \Delta_{\alpha\beta,kl} &= \int_0^\tau\dd{t_1}\int_0^{t_1}\dd{t_2}\expval{b_\alpha(t_1) b_\beta(t_2)}\ctrlmat_{\alpha k}(t_1)\Rc_{\beta l}(t_2), \label{eq:frequency_shift:time} \\
    \Gamma_{\alpha\beta,kl} &= \int_0^\tau\dd{t_1}\int_0^\tau\dd{t_2}\expval{b_\alpha(t_1) b_\beta(t_2)}\ctrlmat_{\alpha k}(t_1)\Rc_{\beta l}(t_2).  \label{eq:decay_amplitudes:time}
\end{align}
The frequency shifts \freqshifts correspond to the first term in \cref{eq:cumulant:truncated}, hence incurring coherent errors, \ie generalized axis and overrotation errors. They reflect a perturbative correction to the quantum evolution due to a change of the Hamiltonian at two points in time, and thus time ordering matters. Conversely, the decay amplitudes \decayamps correspond to the second term and capture the decoherence. These terms are due to an incoherent average that only takes classical correlations into account, so that time ordering does not play a role. Note that \cref{eq:cumulant:truncated:liouville} together with \cref{eq:cumulant_expansion} constitutes an exact version (in the Liouville representation) of Eq.~(4) from \citer{Cerfontaine2021} for Gaussian noise. The approximation of \citer{Cerfontaine2021} is obtained by expanding the exponential to linear order and neglecting the second order terms \freqshifts.

For a single qubit and \basis the Pauli basis one can make use of the simple commutation relations so that the cumulant function takes the form (see \cref{appsec:derivations:cumulant:pauli})
\begin{equation}\label{eq:cumulant:truncated:liouville:pauli}
    \cumulantfun_{ij}(\tau) = \begin{cases}
        - \sum_{k\neq i}\decayamps_{kk}                         &\qif* i = j,   \\
        - \freqshifts_{ij} + \freqshifts_{ji} + \decayamps_{ij} &\qif* i\neq j,
    \end{cases}
\end{equation}
for $i,j > 0$ and any $\alpha,\beta$. As mentioned in \cref{sec:theory:transfer_matrix} the cases $j = 0$ and $i = 0$ encode trace-preservation and unitality, respectively, and as such $\cumulantfun_{0j} = \cumulantfun_{i0} = 0$ since our model is both trace-preserving and unital.

\subsection{Calculating the decay amplitudes}\label{sec:theory:decay_amplitudes}
In order to evaluate the cumulant function $\cumulantfun(\tau)$ given by \cref{eq:cumulant:truncated:liouville} and thus the transfer matrix $\ev*{\liouvUe(\tau)}$ from \cref{eq:cumulant_expansion} for a given control operation, we solely require the decay amplitudes $\decayamps_{kl}$ and frequency shifts $\freqshifts_{kl}$ since the trace tensor $T_{ijkl}$ depends only on the choice of basis and is therefore trivial (although quite costly for large dimensions, \cf \cref{sec:performance:basis}) to calculate. In this section, we describe simple methods for calculating $\decayamps_{kl}$ using an extension of the filter function formalism developed by \citet{Green2013} that we introduced in \citer{Cerfontaine2021}. The central quantity of interest will be the control matrix that we already introduced above. It relates the interaction picture noise operators to the operator basis and we will compute it in Fourier space in order to identify the cross-correlation functions with the noise spectral density in \cref{eq:decay_amplitudes:time}. We distinguish between a sequence of quantum gates, as already presented in our related work \cite{Cerfontaine2021}, and a single gate. In the first case the control matrix of the entire sequence can be calculated from those of the individual gates, greatly simplifying the calculation if the latter have been precomputed. This approach gives rise to correlation terms in the expression for $\decayamps_{kl}$ that capture the effects of sequencing gates. In the second case, as was shown by \citet{Green2013}, one can calculate the control matrix for arbitrary single pulses under the assumption of piecewise constant control and we lay out how to adapt the approach for numerical applications.

We start by noting that, because we assumed the noise fields $b_\alpha(t)$ to be wide-sense stationary, that is to say the cross-correlation functions evaluated at two different points in time $t_1$ and $t_2$ depend only on their difference $t_1 - t_2$, we can define the two-sided noise power spectral density $S_{\alpha\beta}(\omega)$ as the Fourier transform of the cross-correlation functions $\expval{b_\alpha(t_1) b_\beta(t_2)}$,
\begin{equation}\label{eq:spectral_density}
    \expval{b_\alpha(t_1) b_{\beta}(t_2)} = \int_{-\infty}^{\infty}\frac{\dd{\omega}}{2\pi} S_{\alpha\beta}(\omega)\e^{-\i\omega (t_1 - t_2)}.
\end{equation}
Note that the spectrum only characterizes the noise fully in the case of Gaussian noise. For non-Gaussian components in the noise, additional polyspectra have in principle to be considered for higher-order correlation functions \cite{Norris2016}. However, since we only discuss second-order contributions which involve two-point correlation functions here, we only need to take $S_{\alpha\beta}(\omega)$ into account. Inserting the definition of the spectral density into \cref{eq:decay_amplitudes:time}, one finds
\begin{equation}\label{eq:decay_amplitudes:freq}
    \decayamps_{\alpha\beta,kl} = \int_{-\infty}^{\infty}\frac{\dd{\omega}}{2\pi}\ctrlmat^\ast_{\alpha k}(\omega)S_{\alpha\beta}(\omega)\ctrlmat_{\beta l}(\omega)
\end{equation}
with $\ctrlmat(\omega) = \int_0^\tau\dd{t}\ctrlmat(t)\e^{\i\omega t}$ the frequency-domain control matrix. Note that $\ctrlmat^\ast(\omega) = \ctrlmat(-\omega)$ because $\ctrlmat(t)$ is real. In the above equation, the fourth order tensor
\begin{equation}\label{eq:filter_function:generalized}
    F_{\alpha\beta,kl}(\omega)\coloneqq\ctrlmat^\ast_{\alpha k}(\omega)\ctrlmat_{\beta l}(\omega)
\end{equation}
is the generalized filter function that captures the susceptibility of the decay amplitudes to noise at frequency $\omega$. For $\alpha = \beta, k = l$, and by summing over the basis elements,
\begin{equation}\label{eq:filter_function:fidelity}
    F_{\alpha}(\omega) = \sum_k\bigl\lvert\ctrlmat_{\alpha k}(\omega)\bigr\rvert^2 = \tr(\Bat\ad(\omega)\Bat(\omega)),
\end{equation}
and this tensor reduces to the canonical \emph{fidelity} filter function \cite{Green2012} from which the entanglement fidelity can be obtained, see \cref{sec:theory:derived_quantities:entanglement_fidelity}. Thus, if the frequency-domain control matrix $\ctrlmat_{\alpha k}(\omega)$ for noise source $\alpha$ and basis element $k$ is known, the transfer matrix can be evaluated by integrating \cref{eq:decay_amplitudes:freq}. Moreover, one can study the contributions of each pair of noise sources $(\alpha, \beta)$ both separately or, at virtually no additional cost and to leading order, collectively by summing over them, $\decayamps_{kl} = \sum_{\alpha\beta}\decayamps_{\alpha\beta,kl}$.

We now discuss how to calculate the control matrix $\ctrlmat(\omega)$ in frequency space for a given control operation. We focus first on sequences of quantum gates, assuming that the control matrices $\ctrlmat\gth{g}(\omega)$ for each gate $g$ have been calculated before.

\subsubsection{Control matrix of a gate sequence}\label{sec:theory:control_matrix:sequence}
For a sequence of gates with precomputed interaction picture noise operators the approach developed by \citet{Green2013} based on piecewise constant control can be adapted to yield an analytical expression for those of the composite gate sequence that is computationally efficient to evaluate \cite{Cerfontaine2021}. Here we review these results to give a complete picture of the formalism. While our results are general and apply to any superoperator representation, we employ the Liouville representation here for its simple composition operation: matrix multiplication. Computationally, this is not the most efficient choice since transfer matrices have dimension $d^2\times d^2$ and thus their matrix multiplication scales unfavorably compared to, for example, left-right conjugation by unitaries (\cf \cref{sec:performance:complexity}). However, because the structure of the control matrix \ctrlmat is similar to that of a transfer matrix (remember that it corresponds to a basis expansion of the interaction picture noise operators), we will obtain a particularly concise expression for the sequence in the following. For a perhaps more intuitive description employing exclusively conjugation by unitaries, we refer the reader to our accompanying publication \citer{Cerfontaine2021}.

\begin{figure}
    \includegraphics{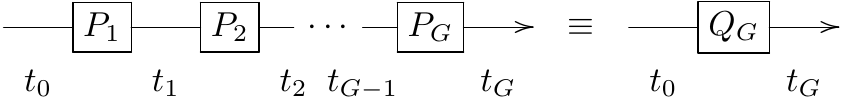}
    \caption{Illustration of a sequence of $G$ gates. Individual gates with propagators $P_g$ start at time $t_{g-1}$ and complete at time $t_g$. The total action from $t_0$ to $t_g$ is given by $Q_g$.}
    \label{fig:gatesequence}
\end{figure}

A sequence of $G$ gates with propagators $P_g = \Uc(t_g, t_{g-1}), g\in\lbrace 1,\dotsc,G\rbrace$ that act during the $g$-th time interval $(t_{g-1}, t_g]$ with $t_0 =  0, t_G = \tau$ as illustrated in \cref{fig:gatesequence} is considered. The cumulative propagator of the sequence up to time $t_g$ is then given by $Q_g = \prod_{g'=g}^0 P_{g'}$ with $P_0 = \eye$ and its Liouville representation denoted by $\liouvQ\gth{g}$. Furthermore, the control matrix of the $g$-th pulse at the time $t - t_{g-1}$ relative to the start of segment $g$ is
\begin{equation}\label{eq:control_matrix:pulse:time}
    \ctrlmat\gth{g}_{\alpha k}(t - t_{g-1}) = \tr(\Uc\ad(t, t_{g-1})B_\alpha(t - t_{g-1}) \Uc(t, t_{g-1}) C_k).
\end{equation}
We can now exploit the fact that in the transfer matrix picture quantum operations compose by matrix multiplication to write the total control matrix at time $t\in (t_{g-1}, t_g]$ as
\begin{equation}\label{eq:control_matrix:sequence:time}
    \ctrlmat(t) = \ctrlmat\gth{g}(t - t_{g-1})\liouvQ\gth{g-1}.
\end{equation}
The Fourier transform of \cref{eq:control_matrix:sequence:time} can then be obtained by evaluating the transform of each gate separately,
\begin{gather}
    \ctrlmat(\omega) = \sum_{g = 1}^G \e^{\i\omega t_{g-1}}\ctrlmat\gth{g}(\omega)\liouvQ\gth{g-1} \label{eq:control_matrix:sequence:freq}\\
    \ctrlmat\gth{g}(\omega) = \int_0^{\Delta t_g}\dd{t}\e^{\i\omega t}\ctrlmat\gth{g}(t), \label{eq:control_matrix:pulse:freq}
\end{gather}
with $\Delta t_g = t_g - t_{g-1}$ the duration of gate $g$. Hence, calculating the control matrix of the full sequence requires only the knowledge of the temporal positions, encoded in the phase factors $\e^{\i\omega t_{g-1}}$, and the total intended action $\liouvQ\gth{g-1}$ of the individual pulses if their control matrices have been precomputed. The sequence structure can thus be exploited to one's benefit. If the same gates appear multiple times during the sequence one can reuse control matrices for equal pulses to facilitate calculating filter functions for complex sequences with modest computational effort. Most importantly, \cref{eq:control_matrix:sequence:freq} is independent of the inner structure of the individual pulses and therefore takes the same time to evaluate whether they are highly complex or very simple. In \cref{sec:performance:complexity}, we will analyze the computational efficiency of capitalizing on this feature in more detail.

As we have seen, the total control matrix of a composite pulse sequence is given by a sum over the individual control matrices. Since $\ctrlmat(\omega)$ enters \cref{eq:decay_amplitudes:freq} twice, this leads to correlation terms between two gates at different positions in the sequence when computing the total decay amplitudes $\decayamps_{\alpha\beta,kl}$. Inserting \cref{eq:control_matrix:sequence:freq} into \cref{eq:decay_amplitudes:freq} gives
\begin{widetext}
\begin{equation}\label{eq:decay_amplitudes:pulse_correlation}
\begin{split}
    \decayamps_{\alpha\beta,kl} &= \sum_{g,g'=1}^{G}\int_{-\infty}^\infty\frac{\dd{\omega}}{2\pi}
                                     \bigl[\liouvQ^{(g'-1)\dagger}\ctrlmat^{(g')\dagger}(\omega)\bigr]_{k\alpha}
                                     S_{\alpha\beta}(\omega)
                                     \bigl[\ctrlmat\gth{g}(\omega)\liouvQ\gth{g-1}\bigr]_{\beta l}
                                     \e^{\i\omega(t_{g-1} - t_{g'-1})} \\
                                &\eqqcolon\sum_{g,g'=1}^{G}\int_{-\infty}^\infty\frac{\dd{\omega}}{2\pi}
                                     S_{\alpha\beta}(\omega) F_{\alpha\beta,kl}\gth{gg'}(\omega)
\end{split}
\end{equation}
\end{widetext}
where we defined the pulse correlation filter function $F_{\alpha\beta,kl}\gth{gg'}(\omega)$ that captures the temporal correlations between pulses at different positions $g$ and $g'$ in the sequence. Unlike regular filter functions, these can be negative for $g\neq g'$ and therefore reduce the overall noise susceptibility of a sequence given by $F(\omega) = \sum_{gg'}F\gth{gg'}(\omega)$. We have thus gained a concise description of the noise-cancelling properties of gate sequences: in this picture, they arise purely from the concatenation of different pulses, quantifying, for instance, the effectiveness of dynamical decoupling (DD) sequences \cite{Cerfontaine2021}.

\subsubsection{Control matrix of a single gate}\label{sec:theory:control_matrix:pulse}
Previous efforts have derived the control matrix analytically for selected pulses such as dynamical decoupling (DD) sequences \cite{Cywinski2008}, special dynamically corrected gates (DCGs) \cite{Gungordu2018}, as well as developed a general analytic framework \cite{Green2012,Green2013}. However, analytical solutions might not always be accessible, \eg for numerically optimized pulses, and are generally laborious to obtain. Therefore, we now detail a method to obtain the control matrix numerically under the assumption of piecewise constant control. Our method is similar in spirit to that of \citet{Green2012} for single qubits with $d=2$, but whereas those authors computed analytical solutions to the relevant integrals during each time step, here we use matrix diagonalization to obtain the propagator of a control operation to make the approach amenable to numerical implementation. This allows carrying out the Fourier transform of the control matrix \cref{eq:control_matrix} analytically by writing the control propagators in terms of their eigenvalues in diagonal form.

We divide the total duration of the control operation, $\tau$, into $G$ intervals $(t_{g-1}, t_{g}]$ of duration $\Delta t_g$ with $g\in\lbrace 0,\dotsc,G\rbrace$ and $t_0 =  0, t_G = \tau$. We then approximate the control Hamiltonian as constant within each interval so that within the $g$-th
\begin{align}\label{eq:hamiltonian:control:piecewise}
    \Hc(t) = \Hc\gth{g} = \mr{const.}
\end{align}
and similarly the deterministic time dependence of the noise operators as $\Ba(t) = s_\alpha(t)\Ba = s_\alpha\gth{g}\Ba$. Under this approximation we can diagonalize the time-independent Hamiltonians $\Hc\gth{g}$ with eigenvalues $\omega_i\gth{g}$ numerically and write the time evolution operator that solves the noise-free Schr\"odinger equation as $\Uc(t, t_{g-1}) = V\gth{g}D\gth{g}(t, t_{g-1})V^{(g)\dagger}$. Here, $V\gth{g}$ is the unitary matrix of eigenvectors of $\Hc\gth{g}$ and the diagonal matrix $D_{ij}\gth{g}(t, t_{g-1}) = \delta_{ij}\exp\{-\i\omega_i\gth{g}(t - t_{g-1})\}$ contains the time evolution of the eigevalues. Using this result together with $Q_{g-1}$, the cumulative propagator up to time $t_{g-1}$, we can acquire the total time evolution operator at time $t$ from $\Uc(t) = \Uc(t,0) = \Uc(t, t_{g-1}) Q_{g-1}$. We then substitute this relation into the definition of the control matrix, \cref{eq:control_matrix}, and obtain
\begin{align}
    \ctrlmat_{\alpha k}(t) = s_\alpha\gth{g}&\mr{tr}\Bigl(Q_{g-1}\ad V\gth{g} D^{(g)\dagger}(t, t_{g-1}) V^{(g)\dagger} B_\alpha \notag\\
                                            &      \times\,V\gth{g} D\gth{g}(t, t_{g-1})V^{(g)\dagger} Q_{g-1} C_k\Bigr) \\
                   \eqqcolon s_\alpha\gth{g}&\sum_{ij}\bar{B}_{\alpha,ij}\gth{g}\bar{C}_{k,ji}\gth{g}
                                   \e^{\i\Omega_{ij}\gth{g}(t - t_{g-1})}, \label{eq:control_matrix:pulse:time:piecewise}
\end{align}
where $\Omega_{ij}\gth{g} = \omega_i\gth{g} - \omega_j\gth{g}$, $\bar{C}_{k}\gth{g} = V^{(g)\dagger}Q_{g-1} C_k Q_{g-1}\ad V\gth{g}$, and $\bar{B}_{\alpha}\gth{g} = V^{(g)\dagger} B_\alpha V\gth{g}$. Carrying out the Fourier transform of \cref{eq:control_matrix:pulse:time:piecewise} to get the frequency-domain control matrix of the pulse generated by the Hamiltonian from \cref{eq:hamiltonian:control:piecewise} is now straightforward since the integrals involved are over simple exponential functions. We obtain
\begin{equation}\label{eq:control_matrix:pulse:freq:calculation}
    \ctrlmat_{\alpha k}(\omega) = \sum_{g = 1}^G s_\alpha\gth{g}\e^{\i\omega t_{g-1}}\tr(\bigl[\bar{B}_\alpha\gth{g}\circ I\gth{g}(\omega)\bigr]\bar{C}_k\gth{g})
\end{equation}
with $I_{ij}\gth{g}(\omega) = -\i(\e^{\i(\omega + \Omega_{ij}\gth{g})\Delta t_g} - 1)/(\omega + \Omega_{ij}\gth{g})$ and the Hadamard product $(A\circ B)_{ij}\coloneqq A_{ij}\cdot B_{ij}$. \Cref{eq:control_matrix:pulse:freq:calculation} is readily evaluated on a computer and thus enables the calculation of filter functions of arbitrary control sequences, either on its own or in conjunction with \cref{eq:control_matrix:sequence:freq}. A similar expression is obtained for representations other than the Liouville representation.

\subsection{Calculating the frequency shifts}\label{sec:theory:frequency_shifts}
The frequency shifts $\freqshifts_{\alpha\beta,kl}$ in \cref{eq:cumulant:truncated:liouville} correspond to the second order of the \gls{me} and thus involve a double integral with a nested time dependence. This makes their evaluation more involved than that of the decay amplitudes $\decayamps_{\alpha\beta,kl}$ and, in contrast to the previous section, we cannot identify a concatenation rule or single out correlation terms as in \cref{eq:decay_amplitudes:pulse_correlation}. However, we can still apply the approximation of piecewise constant control and follow a similar approach as in \cref{sec:theory:control_matrix:pulse} to compute \freqshifts in Fourier space. Since these terms correspond to a coherent gate error that can in principle be calibrated out in experiments we will not go into much detail here.

We follow the arguments made above for the decay amplitudes and express the cross-correlation functions $\ev{b_\alpha(t) b_\beta(t')}$ by their Fourier transform, the spectral density $S_{\alpha\beta}(\omega)$, using \cref{eq:spectral_density}. Inserting this equation into the definition of the frequency shifts in the time domain, \cref{eq:frequency_shift:time}, yields
\begin{multline}\label{eq:frequency_shifts:time}
    \freqshifts_{\alpha\beta,kl} = \int_{-\infty}^\infty\frac{\dd{\omega}}{2\pi} S_{\alpha\beta}(\omega)
                                   \int_0^\tau\dd{t}\ctrlmat_{\alpha k}(t)\e^{-\i\omega t} \\
                                   \times\int_0^t\dd{t'}\ctrlmat_{\beta l}(t')\e^{\i\omega t'}.
\end{multline}
We again assume piecewise constant time segments so that the inner time integral can be split up into a sum of integrals over complete constant segments $(t_{g'-1},t_{g'}]$ as well as a single integral that contains the last, incomplete segment up to time $t$. That is, taking the time $t$ of the outer integral to be within the interval $(t_{g-1}, t_g]$ we perform the replacement
\begin{equation}
    \int_0^t\dd{t'} \rightarrow \sum_{g'=1}^{g-1}\int_{t_{g'-1}}^{t_{g'}}\dd{t'} + \int_{t_{g-1}}^{t}\dd{t'}.
\end{equation}
We have thus divided our task into two: The first term allows, as before in \cref{sec:theory:control_matrix:sequence,sec:theory:control_matrix:pulse}, to identify the Fourier transform of the control matrix during time steps $g'$ and $g$ for both the inner and the outer integral according to \cref{eq:control_matrix:pulse:freq:calculation}. The second term remains a nested double integral, but now the integrand contains only products of complex exponentials because we assume the control to be constant within the limits of integration. As a next step, we also replace the outer time integral by a sum of integrals over single segments, $\int_0^\tau\dd{t}\rightarrow\sum_{g=1}^G\int_{t_{g-1}}^{t_g}\dd{t}$, to obtain
\begin{widetext}
\begin{equation}\label{eq:frequency_shifts:time:substituted}
    \freqshifts_{\alpha\beta,kl} = \int_{-\infty}^\infty\frac{\dd{\omega}}{2\pi} S_{\alpha\beta}(\omega)
            \sum_{g=1}^{G}\int_{t_{g-1}}^{t_{g}}\dd{t}\e^{-\i\omega t}\ctrlmat_{\alpha k}(t)
            \left\lbrace\sum_{g'=1}^{g-1}\int_{t_{g'-1}}^{t_{g'}}\dd{t'} + \int_{t_{g-1}}^{t}\dd{t'}\right\rbrace
            \e^{\i\omega t'}\ctrlmat_{\beta l}(t').
\end{equation}
Before continuing, we ease notation and define $\ctrlmat(\omega) \eqqcolon \sum_g\mc{G}\gth{g}(\omega)$ with $\mc{G}\gth{g}(\omega)$ obtained from \cref{eq:control_matrix:pulse:freq:calculation} and furthermore adopt the Einstein summation convention for the remainder of this section, meaning multiple subscript indices that appear on only one side of an equality are summed over implicitly. We now proceed like in \cref{sec:theory:control_matrix:pulse} and make use of the piecewise constant approximation to diagonalize the control Hamiltonian during each segment. For the nested integrals, we obtain $\ctrlmat_{\alpha k}(t)$ from \cref{eq:control_matrix:pulse:time:piecewise}, whereas the remaining integrals factorize and we can identify the Fourier transformed quantity $\mc{G}\gth{g}(\omega)$. \Cref{eq:frequency_shifts:time:substituted} then becomes
\begin{equation}\label{eq:frequency_shifts:freq}
    \freqshifts_{\alpha\beta,kl} = \int_{-\infty}^\infty\frac{\dd{\omega}}{2\pi} S_{\alpha\beta}(\omega)\sum_{g=1}^G\left[
        \mc{G}_{\alpha k}^{(g)\ast}(\omega)\sum_{g'=1}^{g-1}\mc{G}_{\beta l}\gth{g'}(\omega) + 
        s_\alpha\gth{g}\bar{B}_{\alpha,ij}\gth{g}\bar{C}_{k,ji}\gth{g} I_{ijmn}\gth{g}(\omega)
        \bar{C}_{l,nm}\gth{g}\bar{B}_{\beta,mn}\gth{g} s_\beta\gth{g}
    \right]
\end{equation}
with $\bar{B}_{\alpha,ij}\gth{g},\bar{C}_{k,ij}\gth{g},\Omega_{ij}\gth{g}$ as defined above in \cref{sec:theory:control_matrix:pulse} and
\begin{equation}\label{eq:frequency_shifts:integral}
    I_{ijmn}\gth{g}(\omega) = \int_{t_{g-1}}^{t_g}\dd{t}\e^{\i\Omega_{ij}\gth{g}(t - t_{g-1}) - \i\omega t}
                              \int_{t_{g-1}}^{t}\dd{t'}\e^{\i\Omega_{mn}\gth{g}(t' - t_{g-1}) + \i\omega t'}.
\end{equation}
\end{widetext}
Explicit results for the integration in \cref{eq:frequency_shifts:integral} are given in \cref{appsec:derivations:frequency_shifts:integral}. To calculate the frequency shifts \freqshifts, we can thus reuse the quantity $\mc{G}\gth{g}(\omega)$ also required for the decay amplitudes \decayamps. The only additional computation, apart from contraction, involves the $G$ integrations $I_{ijmn}\gth{g}(\omega)$. Importantly, \cref{eq:frequency_shifts:freq} has the same structure as the corresponding \cref{eq:decay_amplitudes:freq} for \decayamps in that the individual entries of \freqshifts are given by an integral over the spectral density of the noise multiplied with a -- in this case second order -- filter function that describes the susceptibility to noise at frequency $\omega$:
\begin{equation}\label{eq:frequency_shifts:filter_function}
    \freqshifts_{\alpha\beta,kl} = \int_{-\infty}^\infty\frac{\dd{\omega}}{2\pi} S_{\alpha\beta}(\omega) F_{\alpha\beta,kl}\gth{2}(\omega).
\end{equation}

\subsection{Computing derived quantities}\label{sec:theory:derived_quantities}
By means of \cref{eq:control_matrix:sequence:freq,eq:control_matrix:pulse:freq:calculation,eq:frequency_shifts:freq}, one can obtain the cumulant function $\cumulantfun(\tau)$ from \cref{eq:cumulant:truncated:liouville} and hence the error process $\expval*{\liouvUe(\tau)}$ from \cref{eq:cumulant_expansion} for an arbitrary sequence of gates. From this, several quantities of interest for the characterization of a given control operation can be extracted. We explicitly review the average gate and state fidelities as well as expressions to quantify leakage here, but emphasize that this is not exhaustive. Because for many applications the noise is weak and hence the parameter $\xi\ll 1$, we will in the following expand the exponential in \cref{eq:error_propagator} to leading order in $\xi$ in the following. That is, we approximate (remember that $\cumulantfun(\tau)\in\order{\xi^2})$
\begin{equation}\label{eq:error_transfer_matrix:approx}
\ev*{\liouvUe(\tau)}\approx\eye + \cumulantfun(\tau).
\end{equation}
For Gaussian noise, higher order corrections can be obtained either by explicitly calculating higher powers of \cumulantfun or by numerically evaluating the exponential of the cumulant function. The former method often leads to simpler expressions than \cref{eq:cumulant:truncated:liouville} for which the trace tensor $T_{ijkl}$ need not be computed directly. In the weak-noise regime, one can also define specific filter functions for each derived quantity that are given in terms of linear combinations of the generalized filter functions $F_{\alpha\beta,kl}(\omega)$. The ensemble expectation value of the quantity can then be obtained directly from the overlap with the spectral density, $\int\flatfrac{\dd{\omega}}{2\pi} F(\omega) S(\omega)$. Finally, we will drop the averaging brackets and the argument of the error transfer matrix $\ev*{\liouvUe(\tau)}$ for brevity in the following.

\subsubsection{Average gate and entanglement fidelity}\label{sec:theory:derived_quantities:entanglement_fidelity}
The average gate fidelity is a commonly quoted figure of merit used to characterize physical gate implementations \cite{Loss1998,Ladd2010,Chow2012,Veldhorst2014,Yoneda2018}. It represents the fidelity between an implementation \liouvU and the ideal gate \liouvQ averaged over the uniform Haar measure. Since $\avgfid(\liouvU, \liouvQ) = \avgfid(\liouvQ\ad\circ\liouvU, \eye) = \avgfid(\liouvUe)$, the average gate fidelity can be obtained from the error channel \liouvUe as \cite{Horodecki1999,Nielsen2002}
\begin{align}
    \avgfid(\liouvUe) &= \frac{\tr\liouvUe + d}{d(d+1)} \label{eq:fidelity:avg}\\
                      &= \frac{d\times\entfid(\liouvUe) + 1}{d + 1}, \label{eq:fidelity:avg-ent}
\end{align}
where $d$ is the system dimension and $\entfid(\liouvUe) = \tr\liouvUe / d^2$ is the entanglement fidelity. In the low-noise regime where \cref{eq:error_transfer_matrix:approx} holds, we can write the entanglement fidelity in terms of the cumulant function $\cumulantfun_{\alpha\beta}$ approximately as
\begin{align}\label{eq:fidelity:ent}
    \entfid(\liouvUe) &= 1 + \frac{1}{d^2}\sum_{\alpha\beta}\tr\cumulantfun_{\alpha\beta} \\
                      &\eqqcolon 1 - \sum_{\alpha\beta}\infid_{\alpha\beta}(\liouvUe).
\end{align}
Here, we defined $\infid_{\alpha\beta}$, the infidelity due to a pair of noise sources $(\alpha,\beta)$. As we show in \cref{appsec:derivations:fidelity}, we can simplify the trace of the cumulant function so that the infidelity reads
\begin{equation}\label{eq:infidelity:ent}
    \infid_{\alpha\beta} = \frac{1}{d}\tr\decayamps_{\alpha\beta}.
\end{equation}
\Cref{eq:infidelity:ent} reduces to Eq.~(32) from \citer{Green2012} for a single qubit ($d=2$) and pure dephasing noise up to a different normalization convention; by pulling the trace through to the generalized filter function $F_{\alpha\beta,kl}(\omega)$ in \cref{eq:decay_amplitudes:freq}, we recover the relation (setting $\alpha=\beta$ for simplicity)
\begin{equation}\label{eq:infidelity:ent:integral}
    \infid_{\alpha} = \frac{1}{d}\int_{-\infty}^\infty\frac{\dd{\omega}}{2\pi} S_{\alpha}(\omega) F_{\alpha}(\omega)
\end{equation}
with the fidelity filter function $F_{\alpha}(\omega)$ given by \cref{eq:filter_function:fidelity}. Notably, only the decay amplitudes \decayamps contribute to the fidelity to leading order since the frequency shifts \freqshifts are antisymmetric and therefore vanish under the trace.

\subsubsection{State fidelity and measurements}\label{sec:theory:derived_quantities:state_fidelity-measurements}
In the context of quantum information processing we are often interested in the probability of measuring the expected state during readout. We can extract this projective readout probability from the transfer matrix in \cref{eq:error_propagator} by inspecting the transition probability, or state fidelity, between a pure state $\rho = \op{\psi}$ and an arbitrary state $\sigma$ that evolves according to the quantum operation $\qp: \sigma\rightarrow\qp(\sigma)$. Using the double braket notation introduced at the beginning of \cref{sec:theory} we then define the state fidelity as
\begin{equation}\label{eq:fidelity:state}
\begin{split}
    \fid(\kpsi, \liouvU(\sigma)) &= \tr(\rho\qp(\sigma)) \\
                               &= \dip{\rho}{\qp(\sigma)} \\
                               & =  \dmel{\rho}{\qp}{\sigma},
\end{split}
\end{equation}
where we have expressed the density matrices by vectors on the Liouville space \Lspace and \qp as a transfer matrix. We can thus calculate arbitrary pure state fidelities by simple matrix-vector multiplications of the transfer matrices $\qp = \liouvQ\liouvUe$ and the vectorized density matrices $\dket{\rho}$ and $\dket{\sigma}$. In \cref{sec:examples:randomized_benchmarking} we employ this measure to simulate a \gls{rb} experiment where return probabilities are of interest so that $\fid(\kpsi, \qp(\rho)) = \dbra{\rho}\liouvQ\liouvUe\dket{\rho}$.

General measurements can be incorporated in the superoperator formalism we have employed here in a straightforward manner using the \gls{povm} formalism \cite{Wallman2014,Greenbaum2015}. \Glspl{povm} constitute a set of Hermitian, positive semidefinite operators $\lbrace E_i\rbrace_i$ (in contrast to the projective measurement $\lbrace\op{\psi}{\psi},\eye - \op{\psi}{\psi}\rbrace$) that fulfill the completeness relation $\sum_i E_i = \eye$ and in the double braket notation may be represented as the row vectors $\lbrace\dbra{E_i}\rbrace_i$ in Liouville space. Consequently, the measurement probability for outcome $E_i$ is given by $\dip{E_i}{\qp(\sigma)} = \dmel{E_i}{\qp}{\sigma}$ if the system was prepared in the state $\sigma$ and evolved according to \qp.

\subsubsection{Leakage}\label{sec:theory:derived_quantities:leakage}
In many physical implementations qubits are not encoded in real two-level systems but in two levels of a larger Hilbert space (\eg transmon \cite{Koch2007} or singlet-triplet \cite{Petta2005} spin qubits) such that population can leak between this computational subspace and other energy levels. Thus, it is often of interest to quantify leakage when assessing gate performance. Recently, \citet{Wood2018} have suggested two separate measures for quantifying leakage out of the computational subspace on the one hand and seepage into the subspace on the other. With the filter function formalism and the transfer matrix of the error process given by \cref{eq:error_propagator,eq:cumulant:truncated:liouville}, we can easily extract these quantities.

Using the definitions from \citer{Wood2018} and the double braket notation we can write the leakage rate generated by a quantum operation \qp as
\begin{subequations}
\begin{equation}\label{eq:leakage}
    \leak_c(\qp)\coloneqq\frac{1}{d_c}\dbra{\Pi_\ell}\qp\dket{\Pi_c}
\end{equation}
and the seepage rate as
\begin{equation}\label{eq:seepage}
    \leak_\ell(\qp)\coloneqq\frac{1}{d_\ell}\dbra{\Pi_c}\qp\dket{\Pi_\ell}.
\end{equation}
\end{subequations}
Here, $\Pi_{c,\ell}$ are projectors onto the computational and leakage subspaces, respectively, and $d_{c,\ell}$ the corresponding dimensions. For unital channels the leakage and seepage rates are not independent but satisfy $d_c \leak_c = d_\ell \leak_\ell$ \cite{Wood2018} so that we only need to consider one of the above expressions here (\cf \cref{sec:theory:transfer_matrix:derivation}).

\Cref{eq:leakage,eq:seepage} can be used to determine both coherent and incoherent leakage separately by substituting \liouvQ or \liouvUe, respectively, for \qp. While the former is due to systematic errors of the applied pulse and could thus be corrected for by calibration, the latter is induced by noise only. Alternatively, the leakage from both contributions can also be determined collectively by substituting \liouvU for \qp.

\section{Performance analysis and efficiency improvements}\label{sec:performance}
In this section we focus on computational aspects of the formalism, remarking first on several mathematical simplifications that make the calculation of control matrices and decay amplitudes more economical. Following this, we investigate the computational complexity of the method in comparison with Monte Carlo techniques and show that our software implementation surpasses the latter's performance in relevant parameter regimes.

\subsection{Periodic Hamiltonians}\label{sec:performance:periodic_hamiltonians}
If the control Hamiltonian is periodic, that is $\Hc(t) = \Hc(t + T)$, we can reduce the computational effort of calculating the control matrix by potentially orders of magnitude (see \cref{sec:examples:rabi_driving} for an application in Rabi driving). We start by making the following observations: First, the frequency domain control matrix of every period of the control is the same so that $\ctrlmat\gth{g}(\omega) = \ctrlmat\gth{1}(\omega)$. Moreover, $\e^{\i\omega\Delta t_g} = \e^{\i\omega T}$ for all $g$ so that $\e^{\i\omega t_{g-1}} = \e^{\i\omega T(g - 1)}$ and by the composition property of transfer matrices $\liouvQ\gth{g-1} = \left[\liouvQ\gth{1}\right]^{g-1}$ where the superscript without parentheses denotes matrix power. We can then simplify \cref{eq:control_matrix:sequence:freq} to read
\begin{equation}\label{eq:control_matrix:sequence:periodic:explicit}
    \ctrlmat(\omega) = \ctrlmat\gth{1}(\omega)\sum_{g=0}^{G-1}\left[\e^{\i\omega T}\liouvQ\gth{1}\right]^g.
\end{equation}
Furthermore, if the matrix $\eye - \e^{\i\omega T}\liouvQ\gth{1}$ is invertible, which is typically the case for the vast majority of values of $\omega$, the previous expression can be rewritten as
\begin{equation}\label{eq:control_matrix:sequence:periodic:simplified}
    \ctrlmat(\omega) = \ctrlmat\gth{1}(\omega)\Bigl(\eye - \e^{\i\omega T}\liouvQ\gth{1}\Bigr)^{-1}
        \Bigl(\eye - \bigl[\e^{\i\omega T}\liouvQ\gth{1}\bigr]^G\Bigr)
\end{equation}
by evaluating the sum as a finite Neumann series. \Cref{eq:control_matrix:sequence:periodic:simplified} offers a significant performance benefit over regular concatenation in the case of many periods $G$ as we will show in \cref{sec:performance:complexity}. Beyond numerical advantages, it also provides an analytic method for studying filter functions of periodic driving Hamiltonians.

\subsection{Extending Hilbert spaces}\label{sec:performance:extending_hilbert_spaces}
Examining \cref{eq:control_matrix}, we can see that the columns of the control matrix and therefore also the filter function are invariant (up to normalization) under an extension of the Hilbert space. This allows parallelizing pulses with precomputed control matrices in a very resource-efficient manner if one chooses a suitable operator basis. Note that the same also applies to other representations of quantum operations.

Suppose we extend the Hilbert space $\Hspace_1$ of a gate for which we have already computed the control matrix by a second Hilbert space $\Hspace_2$ so that $\Hspace_{12} = \Hspace_1\otimes\Hspace_2$. If we can find an operator basis whose elements separate into tensor products themselves, \ie $\basis_{12} = \basis_1\otimes\basis_2$ as for the Pauli basis (\cf \cref{sec:performance:basis}), the control matrix of the composite gate defined on $\Hspace_{12}$ has the same non-trivial columns as that of the original gate on $\Hspace_1$ up to a different normalization factor. The remaining columns are simply zero. This is because the trace over a tensor product factors into traces over the individual subsystems so that $\ctrlmat_{\alpha k}(t)\propto\mr{tr}\bigl([U_1\ad\otimes U_2\ad][B_\alpha\otimes\eye][U_1\otimes U_2][\eye\otimes C_k]\bigr) = \mr{tr}\bigl(U_1\ad B_\alpha U_1\eye\bigr)\mr{tr}\bigl(U_2\ad\eye U_2 C_k\bigr) = 0$ since we assumed that the noise operators $B_\alpha$ are traceless (\cf \cref{sec:theory:transfer_matrix:derivation}).

Generalizing this result to multiple originally disjoint Hilbert spaces we write the composite space as $\Hspace = \bigotimes_i\Hspace_i$ and the corresponding basis as $\basis = \bigotimes_i\basis_i$. The control matrix of the composite pulse on \Hspace is then a combination of the columns of the control matrices on $\Hspace_i$ for noise operators $B_\alpha$ that are non-trivial, \ie not the identity, only on their original space. For noise operators defined on more than one subspace, \eg $B_{ij} = B_i\otimes B_j, B_i\in\Hspace_i, B_j\in\Hspace_j$, this does not hold anymore and the corresponding row in the composite control matrix needs to be computed from scratch.

One can thus reuse precomputed control matrices beyond the concatenation laid out above when studying multi-qubit pulses or algorithms. For concreteness, consider a set of one- and two-qubit pulses whose control matrices have been precomputed. We can then remap those control matrices to any other qubit in a larger register if the entire Hilbert space is defined by the tensor product of the single-qubit Hilbert spaces, and even map the control matrices of two different pulses to the same time slot on different qubits. Thus, we do not need to perform the possibly costly computation of the control matrices again but instead only need to remap the columns of \ctrlmat to the equivalent basis elements in the basis of the complete Hilbert space, making the assembly of algorithms that consist of a limited set of gates which are used at several points in the algorithm more efficient. In \cref{sec:examples:qft} we simulate a four-qubit \gls{qft} algorithm making use of the shortcuts described here.

\subsection{Operator bases}\label{sec:performance:basis}
Up to this point, we have not explicitly specified the basis that defines the Liouville representation. The only conditions imposed by \cref{eq:basis} are orthonormality with respect to the Hilbert-Schmidt product and that the basis elements are Hermitian. Yet, the choice of operator basis can have a large impact on the time it takes to compute the control matrix as discussed in the previous section. We therefore give a short overview over two possible choices in the following. As we are mostly interested in the computational properties, we represent linear operators in $\mathsf{L}(\Hspace)$ as matrices on $\mathbb{C}^{d\times d}$.

The $n$-qubit Pauli basis fulfills the requirements set by \cref{eq:basis} and furthermore allows for the simplifications described before. In our normalization convention where $\dotHS{C_i}{C_i} = \eye$ it can be written as
\begin{equation}\label{eq:basis:pauli}
    \left\lbrace\sigma_i\right\rbrace_{i=0}^{d^2-1} = \left\lbrace
                                                          \frac{\eye}{\sqrt{2}},
                                                          \frac{\px}{\sqrt{2}},
                                                          \frac{\py}{\sqrt{2}},
                                                          \frac{\pz}{\sqrt{2}}
                                                      \right\rbrace^{\otimes n}
\end{equation}
with the Pauli matrices $\px,\py$ and $\pz$ . While it is obvious that it is separable, meaning it factors into tensor products of the single-qubit Pauli matrices, the dimension of the Pauli basis is restricted to powers of two, \ie $d = 2^n$. An operator basis without this restriction is the generalized Gell-Mann (GGM) basis \cite{Kimura2003,Bertlmann2008}. In the following we will discuss optimizations pertaining to this basis that are also implemented in the software (see \cref{sec:software}).

The GGM matrices are a generalization of the Gell-Mann matrices known from particle physics to arbitrary dimensions. In our normalization convention, the basis (excluding the identity element) is given by \cite{Hioe1981}
\begin{subequations}\label{eq:basis:ggm}
\begin{equation}\tag{\ref{eq:basis:ggm}}
    \left\lbrace\Lambda_i\right\rbrace_{i=1}^{d^2-1} = \left\lbrace u_{jk}, v_{jk}, w_{l}\right\rbrace_{j,k,l}
\end{equation}
with
\begin{align}
    u_{jk} &= \frac{1}{\sqrt{2}}\left(\dyad{j}{k} + \dyad{k}{j}\right), \label{eq:basis:ggm:u}\\
    v_{jk} &= -\frac{\i}{\sqrt{2}}\left(\dyad{j}{k} - \dyad{k}{j}\right), \label{eq:basis:ggm:v}\\
    w_{l} &= \frac{1}{\sqrt{l(l+1)}}\left(\sum_{m=1}^l\dyad{m}{m} - l\dyad{l+1}{l+1}\right), \label{eq:basis:ggm:w}
\end{align}
\end{subequations}
for $1\leq j < k\leq d$, $1\leq l\leq d - 1$, and an orthonormal vector basis $\lbrace\ket{j}\rbrace_{j=1}^d$ of the Hilbert space. Expanding an arbitrary matrix $A\in\mathbb{C}^{d\times d}$ in the basis of \cref{eq:basis:ggm} is then simply a matter of adding up the corresponding matrix elements of $A$ according to \cref{eq:basis:ggm:u,eq:basis:ggm:v,eq:basis:ggm:w}. For instance, the expansion coefficient for the first symmetric basis element is given by $u_{12} = \flatfrac{(A_{12}\dyad{1}{2} + A_{21}\dyad{2}{1})}{\sqrt{2}}$. The explicit construction prescription of the GGM basis thus allows calculating inner products of the form $\ip{\Lambda_j}{A}$ at constant cost instead of the quadratic cost of the trace of a matrix product, speeding up the computation of the transfer matrix from \cref{eq:liouville_representation} (in which case $A = \qp(\Lambda_k)$). In numerical experiments, calculating the transfer matrix of a unitary $U$ with dimension $d$ and precomputed matrix products $A_k  =  U \Lambda_k U\ad$ scaled as $\sim d^{4.16}$. This agrees with the expected scaling of $\sim d^4$ (a transfer matrix has $d^2\times d^2$ elements) and presents a significant improvement over the explicit calculation with trace overlaps $\tr(\Lambda_j A_k)$ that we observed to scale as $\sim d^{5.93}$ (we expected $\sim d^6$).

Further inspection of the GGM basis additionally reveals an increasing sparsity for large $d$ (the filling factor scales roughly with $d^{-2}$), so that it is well suited for computing the trace tensor \cref{eq:trace_tensor}. Since this tensor has $d^8$ elements, the amount of memory required for a dense representation becomes unreasonably large quite quickly. To overcome this constraint, we can use a GGM basis instead of a dense basis like the Pauli basis (which has a filling factor of $\flatfrac{1}{2}$). In this case, the resulting tensor is also sparse because the overlap between different basis elements is small. This not only enables storing the tensor in memory but also makes the calculation much faster since one can employ algorithms optimized for operations on sparse arrays (see \cref{sec:software}).

As an illustration, consider a system of four qubits so that the Hilbert space has dimension $d = 2^4$. An operator basis for this space has $d^2 = 2^8$ elements and consequently the tensor $T_{ijkl}$ has $(2^8)^4 = 2^{32}$ entries. Using \SI{128}{\bit} complex floats to represent the entries the tensor would take up $\approx\SI{68}{\giga\byte}$ of memory in a dense format. Conversely, for a GGM basis stored in a sparse data structure, the resulting trace tensor only takes up $\approx\SI{100}{\mega\byte}$ of memory. Furthermore, calculating $T$ takes $\approx\SI{2.89}{\second}$ on an \fastprocessor since a GGM has a low filling factor. By contrast, the same calculation with a Pauli basis takes $\approx\SI{217}{\second}$. This is due to the larger filling factor on the one hand and because sparse matrix multiplication algorithms perform poorly with dense matrices on the other.

\subsection{Computational complexity}\label{sec:performance:complexity}
In order to assess the performance of filter functions (FF) for computing fidelities compared to Monte Carlo (MC) methods, we determine each method's asymptotic scaling behavior as a function of the system dimension $d$. For the filter functions, we calculate the fidelities using \cref{eq:infidelity:ent:integral,eq:filter_function:fidelity} in our software implementation, described in more detail in \cref{sec:software} and hence neglect contributions of $\order{\xi^4}$ from the frequency shifts \freqshifts. Additionally, we distinguish between three different approaches for calculating the control matrix; first, for a single pulse following \cref{eq:control_matrix:pulse:freq:calculation}, second for an arbitrary sequence of pulses following \cref{eq:control_matrix:sequence:freq}, and third for a periodic sequence of pulses following \cref{eq:control_matrix:sequence:periodic:simplified}. For the single pulse, we run benchmarks using exemplary values for the various parameters on a machine with an \slowprocessor and \SI{24}{\giga\byte} of memory. We also discuss the filter function method using left-right conjugation by unitaries instead of the Liouville representation. The latter has higher memory requirements and is expected to perform poorly for large system dimensions $d$ since one deals with $d^2\times d^2$ transfer matrices on a Liouville space \Lspace instead of $d\times d$ unitaries on a Hilbert space \Hspace. In the software package, the calculations are currently implemented in Liouville space and calculation by conjugation is only partially supported through the low-level API. However, both representations perform similarly for small dimensions as we show below. Note that for a fair performance comparison the different nature of errors needs to be kept in mind. Monte Carlo becomes less costly if larger statistical errors can be tolerated, whereas the filter function formalism is typically limited by higher order errors. For reference, the following considerations are summarized in \cref{tab:complexity} for each approach and a representative set of parameters.

To calculate the fidelity using MC, we generate $n_\mr{MC}$ different noise traces that slice every time step $\Delta t$ of the pulse into $n_\mr{seg} = f_\mr{UV}\Delta t$ segments to appropriately sample the spectral density with $f_{\mr{UV}}$ being the ultraviolet cutoff frequency. In total, there are $n_{\Delta t} n_\mr{MC} n_\mr{seg}$ noise samples for each of which the Hamiltonian is diagonalized, exponentiated, and the resulting propagators multiplied to get the final, noisy unitary. The entanglement fidelity is then obtained by averaging the trace overlap $\flatfrac{\tr(Q\ad U)}{d}$ of ideal and noisy unitary over all noise realizations. Taking the complexity of matrix diagonalization to be $\order{d^3}$ and matrix multiplication to be $\order{d^b}$ with $b = 3$ for a naive algorithm and $b = \num{2.376}$ for the Coppersmith-Winograd algorithm \cite{Coppersmith1990}, we expect MC to scale with the dimension $d$ of the problem as $\sim n_{\Delta t} n_\mr{MC} n_\mr{seg} (d^b + d^3)$. For simplicity, we use a white noise spectrum for which $S(\omega) = \mr{const.}$ but note that sampling arbitrary spectra induces additional overhead for MC, depending on which method is used to generate the noise traces. Typical time-domain methods include the simulation of the underlying physical process (like two-state fluctuators) or the application of an inverse Fourier transform to white noise multiplied by a frequency-domain transfer function.

\begin{table}[tbp]
    \renewcommand\arraystretch{1.25}
    \begin{tabular*}{\columnwidth}{@{\extracolsep{\fill}} lc S[table-format=1.1e1]}
    \toprule
    Method                  & Dominating scaling                                                & {Ex. values}  \\
    \colrule
    MC (\Hspace)            & $n_{\Delta t} n_\mr{MC} n_\mr{seg} (d^b + d^3)$                   & 1.3e8 \\
    FF (\Lspace, explicit)  & $n_{\Delta t} n_\omega n_\alpha d^{4} + n_{\Delta t} d^{b+2}$     & 2.4e7 \\
    FF (\Hspace, explicit)  & $n_{\Delta t} n_\omega n_\alpha (d^{2} + d^{b})$                  & 1.4e7 \\
    FF (\Lspace, concat.)   & $G n_\omega n_\alpha d^{4} + G d^{2b}$                            & 2.4e6 \\
    FF (\Hspace, concat.)   & $G n_\omega n_\alpha d^b$                                         & 7.8e5 \\
    FF (\Lspace, periodic)  & $n_\omega (n_\alpha d^{4} + d^{2b} + d^{2b}\log{G})$              & 1.0e5 \\
    \botrule
    \end{tabular*}
    \caption{Complexity scaling of the three approaches for calculating average gate fidelities discussed in the text. \enquote{FF (explicit)} stands for calculating filter functions from scratch following \cref{eq:control_matrix:pulse:freq:calculation}, \enquote{FF (concat.)} for sequences following \cref{eq:control_matrix:sequence:freq}, and \enquote{FF (periodic)} for periodic Hamiltonians following \cref{eq:control_matrix:sequence:periodic:simplified}. \Hspace and \Lspace designate the vector space on which calculations are performed. Example values for the dominant contributions listed in the table are given for matrix multiplication exponent $b = 2.376$, dimension $d = 2$, number of time steps $n_{\Delta t} = 1000$, and number of gates $G = 100$ (corresponding to a sequence of 100 single-qubit gates with 10 time steps each) with the remaining parameters as in \cref{fig:performance:MC_vs_FF}. For increasing $d$ the computational advantage of FF (\Lspace) over MC diminishes but is conserved for FF (\Hspace).}
    \label{tab:complexity}
\end{table}

By contrast, the computational cost of the filter function formalism as realized by \cref{eq:control_matrix:pulse:freq:calculation} is independent of the form of the spectrum. For this approach we find the leading terms to scale as $\sim n_{\Delta t} n_\omega n_\alpha d^{4} + n_{\Delta t} d^{b+2}$ with $n_\alpha$ the number of noise operators and $n_\omega$ the number of frequency samples. Here, the first term is due to the trace in \cref{eq:control_matrix:pulse:freq:calculation} which boils down to the trace of a matrix product, $\sum_{ij} A_{ij} B_{ji}$, that scales with $d^2$ and is performed once for each of the $d^2$ basis elements, $n_\alpha$ noise operators, $n_{\Delta t}$ time steps, and $n_\omega$ frequency points. The second term is due to the transformation $C_k\rightarrow\bar{C}\gth{g}_k$ which requires multiplication of $d\times d$ matrices for every time step and basis element. As $n_\alpha n_\omega < n_\mr{MC} n_\mr{seg}$ for realistic parameters because the ultraviolet cutoff frequency needs to be chosen sufficiently high and the relative error of the method decreases with $\flatfrac{1}{\sqrt{n_\mr{MC}}}$, we expect that in the case of a single pulse the filter function formalism in Liouville representation should outperform Monte Carlo calculations for reasonably small dimensions $d$. Using left-right conjugation, this advantage should hold also for large $d$. In this case the Hadamard product ($\sim d^2$) as well as matrix multiplication ($\sim d^b$) are carried out for each frequency, noise operator, and time step to calculate the interaction picture noise operators $\Bab(\omega)$. We thus find this method to scale with $\sim n_{\Delta t} n_\omega n_\alpha (d^2 + d^b)$.

\begin{figure}[tbp]
    \centering
    \includegraphics{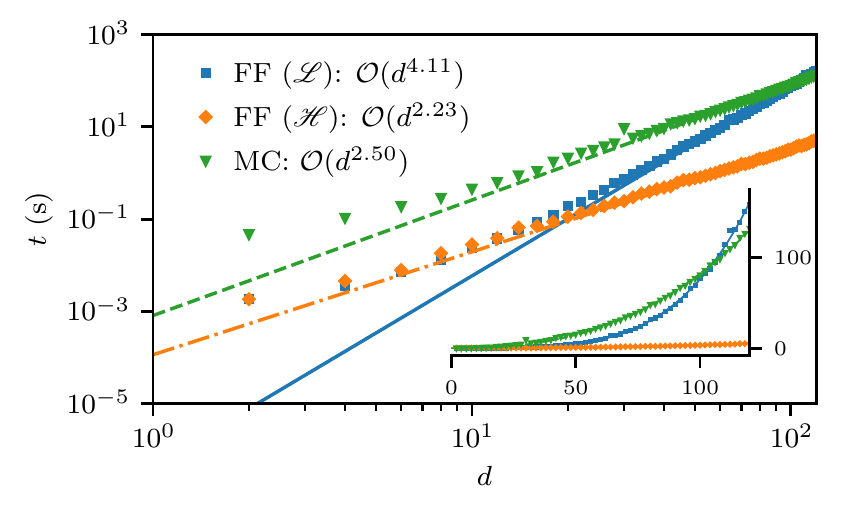}
    \caption{Performance of the formalism using \cref{eq:control_matrix:pulse:freq:calculation} compared to a Monte Carlo method for a single gate as a function of problem dimension $d$. Parameters are: $n_{\Delta t} = \num{1}, n_\alpha = 3, n_\mr{MC} = 100, f_\mr{UV} = \flatfrac{10^2}{\Delta t}, n_\omega = \num{500}$ where $n_\alpha$ is the number of noise operators considered, $n_\mr{MC}$ the number of Monte Carlo trajectories over which is averaged, and $n_\omega$ the number of frequency samples. The calculation using filter functions clearly outperforms MC for small system sizes. For dimensions larger than $d\approx\num{100}$ (roughly equivalent to 7 qubits) Monte Carlo (blue squares) performs better than the \gls{ff} calculation with transfer matrices (green triangles) for this set of parameters and processor due to the better scaling behavior. Using conjugation by unitaries (orange diamonds) significantly outperforms \gls{mc} also for large dimensions. While the fits to $t = a d^b$ (lines) underestimate the leading order exponent due to the data not being in the asymptotic regime, they support the expected relationship of complexity between the approaches. The inset shows the same data on a linear scale, highlighting the different scaling behaviors for large $d$.}
    \label{fig:performance:MC_vs_FF}
\end{figure}

\Cref{fig:performance:MC_vs_FF} shows exemplary wall times for both methods and $d\in[2,120]$ that confirm our expectation. Only for about $d\approx\num{100}$ the overhead from the extra time steps and trajectories over which is averaged is compensated for MC. For smaller dimensions the calculation using FFs is faster by almost two orders of magnitude (see the inset showing the same data in a log-log plot). The lines show fits to $t = a d^b$. The data is not quite in the asymptotic regime due to limited memory so that even for large dimension terms of lower power in $d$ contribute significantly to the run time. Even though this causes the fits to underestimate the exponent $b$, the general trend agrees with our expectation. Note that the crossover does not always occur at the same dimension $d$. On a different system with an \fastprocessor the FF method outperformed MC even for $d = 120$ beyond which available memory limited the simulation.

Quantifying the performance gain from using the control matrices' concatenation property to calculate fidelities of gate sequences is more difficult since it strongly depends on the number of gates occurring multiple times in the sequence (enabling reuse of precomputed control matrices) as well as the complexity of the individual gates. The evaluation using the concatenation rule \cref{eq:control_matrix:sequence:freq} performs asymptotically worse than the evaluation for a complete pulse according to \cref{eq:control_matrix:pulse:freq:calculation} because of higher powers of $d$ dominating the calculation in the former case. Performing the $G$ matrix multiplications $\ctrlmat\gth{g}(\omega)\liouvQ\gth{g-1}$ from \cref{eq:control_matrix:sequence:freq} is of order $\sim G n_\omega n_\alpha d^4$, with $G$ the number of pulses in the sequence. Furthermore, calculating the transfer matrix of the total propagators $Q_{g-1}$ involves multiplication of $d\times d$ matrices for all $d^4$ combinations of basis elements amounting to $\sim G d^{b+4}$. In case the Liouville representation of the individual pulses' total propagators $P_g$, $\liouvP\gth{g}$, have been precomputed, the latter computation can be made more efficient since one can just propagate the transfer matrices $\liouvP\gth{g}$ to obtain the cumulative transfer matrices for the sequence, $\liouvQ\gth{g}=\prod_{g'=g}^0\liouvQ\gth{g'}$, at cost $\sim G d^{2b}$. The restriction to small dimensions does not apply for conjugation by unitaries as in this case the matrix multiplications involve $d\times d$ matrices and we do not have to compute the Liouville representation. We thus obtain a more favorable asymptotic scaling of $\sim G n_\omega n_\alpha d^b$.

Utilizing the concatenation property in the Liouville representation thus corresponds to effectively reducing the number of times the calculations scaling with $\sim n_\omega n_\alpha d^4$ have to be carried out but incurs additional calculations scaling with $\sim d^{2b}$. Accordingly, it provides a performance benefit if a sequence consists of either very complex pulses, in which case single repetitions already make the calculation much more efficient, or of few pulses that occur many times. In the extremal case of $G$ repetitions of a single gate the benefit of employing the concatenation property is most pronounced and can be improved even further utilizing the simplifications laid out in \cref{sec:performance:periodic_hamiltonians}. Since matrix inversion has the same complexity as matrix multiplication and taking a matrix to the $G$-th power requires $\order{\log G}$ matrix multiplications, \cref{eq:control_matrix:sequence:periodic:simplified} should scale with $\sim n_\omega (n_\alpha d^4 + d^{2b} + d^{2b}\log{G})$ (the first two terms are due to the final matrix multiplications and are independent of $G$). It hence allows for a vast speedup over \cref{eq:control_matrix:sequence:freq} in that the asymptotic behavior as a function of the number of gates changes from $\sim G$ to $\sim\log G$. An example of this is presented in \cref{sec:examples:rabi_driving} for the context of Rabi driving. Note that this closed form is a unique feature of the transfer matrix representation and not applicable to conjugation by unitaries.

\section{Software implementation}\label{sec:software}
In this section we give an overview over the \filterfunctions software package implementing the main features of the formalism derived above. This includes the calculation of the decay amplitudes \decayamps and fidelities as well as the calculation of the control matrices for single gates and both generic and periodic sequences of gates. Moreover, control matrices may be efficiently extended to and merged on larger Hilbert spaces. Calculations using unitary conjugation instead of transfer matrices are implemented but at this point not available in the high-level API.

Our software is written in Python and available on GitHub \cite{software} under the GPLv3 license. We also provide a current snapshot in the Supplementary Material \cite{prrSupp}. It features a broad coverage through unit tests and extensive API documentation as well as didactic examples (see \cref{sec:examples}). The package relies on the \numpy \cite{Harris2020} and \scipy \cite{Virtanen2020} libraries for vectorized array operations. Data visualization is handled by \matplotlib \cite{Hunter2007}. For tensor multiplications with optimized contraction order we use \opteinsum \cite{Smith2018} for which \sparse \cite{sparse}, a library aiming to extend the \scipy sparse module to multi-dimensional arrays, serves as a backend in the calculation of the trace tensor from \cref{eq:trace_tensor}. Lastly, the package is written to interface with \qopt \cite{qopt,Teske2021} and \qutip \cite{Johansson2013}, frameworks for the simulation and optimization of open quantum systems, and mirrors the latter's data structure for Hamiltonians ensuring easy interoperability between the two.

\subsection{Package overview}\label{sec:software:overview}
In the \filterfunctions package all operations are understood as sequences of pulses that are applied to a quantum system. These pulses are represented by instances of the \pulsesequence class which holds information about the physical system (control and noise Hamiltonians) as well as the mathematical description (\eg the basis used for the Liouville representation). As indicated above, the Hamiltonians $\Hc(t)$ and $\Hn(t)$ are given in a similar structure as in \qutip. That is, a Hamiltonian is expressed as a sum of Hermitian operators with the time dependence encoded in piecewise constant coefficients so that
\begin{subequations}\label{eq:hamiltonian:software}
\begin{gather}
    \Hc(t) = \sum_i a\gth{g}_i A_i = \mr{const.} \label{eq:hamiltonian:software:control} \\
    \Hn(t) = \sum_\alpha s\gth{g}_\alpha B_\alpha = \mr{const.} \label{eq:hamiltonian:software:noise}
\end{gather}
\end{subequations}
for $t\in (t_{g-1}, t_g], g\in\lbrace 1,\dotsc,G\rbrace$ and where the $a\gth{g}_i$ are the amplitudes of the $i$-th control field. Note that the noise variables $b_\alpha(t)$ are missing from \cref{eq:hamiltonian:software:noise} because they are captured by the spectral density $S(\omega)$. In the software, \cref{eq:hamiltonian:software:control,eq:hamiltonian:software:noise} are represented as lists whose $i$-th element corresponds to a sublist of two elements: the $i$-th operator and the $i$-th coefficients $[a_i\gth{0},\dotsc,a_i\gth{G}]$.

The \pulsesequence class provides methods to calculate and cache the filter function according to \cref{eq:control_matrix:pulse:freq}. Alternatively, filter functions may also be cached manually to permit using the package with analytical solutions for the control matrix. Concatenation of pulses is implemented by the functions \verb|concatenate()| and \verb|concatenate_periodic()| which will attempt to use the cached attributes of the \pulsesequence instances representing the pulses to efficiently calculate the filter function of the composite pulse following \cref{eq:control_matrix:sequence:freq} and \cref{eq:control_matrix:sequence:periodic:simplified}, respectively.

Operator bases fulfilling \cref{eq:basis} are implemented by the \verb|Basis| class. There are two predefined types of bases:
\begin{enumerate}
    \item Pauli bases for $n = 2^d$ qubits from \cref{eq:basis:pauli} and
    \item generalized Gell-Mann (GGM) bases of arbitrary dimension $d$ from \cref{eq:basis:ggm}.
\end{enumerate}
The Pauli basis is both unitary and separable while the GGM basis is sparse for large dimensions but neither unitary nor separable. As mentioned in \cref{sec:performance:extending_hilbert_spaces} (see also \cref{sec:examples:qft}), using a separable basis can provide significant performance benefits for calculating the filter functions of algorithms. On the other hand, a sparse basis makes the calculation of the trace tensor $T_{ijkl}$ and therefore also of the error transfer matrix \liouvUe much faster (\cf \cref{sec:performance:basis}). Additionally, the user can define custom bases using the class constructor.

The error transfer matrix \liouvUe can be calculated for a given pulse and noise spectrum using the \verb|error_transfer_matrix()| function \footnote{Note that while the calculation of the frequency shifts \freqshifts is implemented, it should at time of publication be understood as preliminary and not thoroughly tested}. Various other quantities can be computed from \liouvUe as outlined in \cref{sec:theory:derived_quantities}. Furthermore, the package includes a plotting module that offers several functions, \eg for the visualization of filter functions or the evolution of the Bloch vector using \qutip.

\subsection{Workflow}\label{sec:software:workflow}
We now give a short introduction into the workflow of the \filterfunctions package by showing how to calculate the dephasing filter function of a simple Hahn spin echo sequence \cite{Hahn1950} as an example. The sequence consists of a single $\pi$-pulse of finite duration around the $x$-axis of the Bloch sphere in between two periods of free evolution. We can hence divide the control fields into three constant segments and write the control Hamiltonian as
\begin{equation}
    \Hc^{(\mr{SE})}(t) = \frac{\px}{2}\cdot\begin{cases}
        \flatfrac{\pi}{t_\pi},  &\mr{if\;} \tau\leq t < \tau + t_\pi \\
        0,                      &\mr{otherwise} \\
    \end{cases}
\end{equation}
with $\tau$ the duration of the free evolution period and $t_\pi$ that of the $\pi$ pulse. For the noise Hamiltonian we only need to define the deterministic time dependence $s_\alpha(t)$ and operators $B_\alpha$ since the noise strength is captured by the spectrum $S(\omega)$. Thus we have $s_z(t) =  1$ and $B_z = \flatfrac{\pz}{2}$ for pure dephasing noise that couples linearly to the system.

In the software, we first define a \pulsesequence object representing the spin echo (SE) sequence. As was already mentioned, the control and noise Hamiltonians are given as a list containing lists for every control or noise operator that is considered. These sublists consist of the respective operator as a \numpy array or \qutip \qobj and the amplitudes ($a_i\gth{g}$ or $s_\alpha\gth{g}$) in an iterable data structure such as a list. We can hence instantiate the \pulsesequence with the following code:
\begin{lstlisting}{language=Python}
import filter_functions as ff
import qutip as qt
from math import pi
tau, t_pi = (1, 1e-3)
# Control Hamiltonian for pi rotation in 2nd time step
H_c = [[qt.sigmax()/2, [0, pi/t_pi, 0]]]
# Pure dephasing noise Hamiltonian with linear coupling
H_n = [[qt.sigmaz()/2, [1, 1, 1]]]
# Durations of piecewise constant segments
dt = [tau, t_pi, tau]
ECHO = ff.PulseSequence(H_c, H_n, dt)
\end{lstlisting}
where a basis is automatically chosen since we did not specify it in the constructor in the last line. Calculating the filter function of the pulse for a given frequency vector \verb|omega| can then be achieved by calling
\begin{lstlisting}{language=Python}
F = ECHO.get_filter_function(omega)
\end{lstlisting}
where \verb|F| is the dephasing filter function $F_{zz}(\omega)$ as we only defined a single noise operator. Finally, we calculate the error transfer matrix \liouvUe for the noise spectral density $S_{zz}(\omega) = \omega^{-2}$,
\begin{lstlisting}{language=Python}
S = 1/omega**2
U = ff.error_transfer_matrix(ECHO, S, omega)
\end{lstlisting}
This code uses the control matrix previously cached when the filter function was first computed. Therefore, only the integration in \cref{eq:decay_amplitudes:freq} and the calculation of the trace tensor in \cref{eq:trace_tensor} are carried out in the last line.

An alternative approach to calculate the spin echo filter function is to employ the concatenation property. For this, we interpret the SE as a sequence consisting of three separate pulses. Each of the pulses has a single time segment during which a constant control is applied and concatenating the separate \pulsesequence instances yields the \pulsesequence representing a spin echo. This way analytic control matrices may be used to calculate the control matrix of the composite sequence. Pulses can be concatenated by using either the \verb|concatenate()| function or the overloaded \verb|@| operator:
\begin{lstlisting}{language=Python}
# Define PulseSequence objects as shown above
FID = ff.PulseSequence(...)
NOT = ff.PulseSequence(...)
# Cache the analytic control matrices at frequency omega
FID.cache_control_matrix(omega, B_FID)
NOT.cache_control_matrix(omega, B_NOT)
# Concatenate the pulses
ECHO = FID @ NOT @ FID
\end{lstlisting}
Since we have cached the control matrices of the \texttt{FID} and \texttt{NOT} pulses, that of the \texttt{ECHO} object is also automatically calculated and stored. Concatenating \pulsesequence objects is implemented as an arithmetic operator of the class to reflect the intrinsic composition property of the control matrices.

Further development of the software has focused on making it available in a gate optimization and simulation framework \cite{qopt,Teske2021}. Besides using it to compute decoherence effects and fidelities, analytic derivatives of the filter functions have been implemented to allow for optimizing pulse parameters in the presence of non-Markovian noise within the framework of quantum optimal control \cite{Le2021}.

Additionally, building an interface with \qupulse \cite{qupulse,Humpohl2021}, a software toolkit for parametrizing and sequencing control pulses and relaying them to control hardware, would implement the capability to compute filter functions of pulses assembled in \qupulse, thereby allowing a user in the lab to easily inspect the noise susceptibility characteristics of the pulse they are currently applying to their device.

\section{Example applications}\label{sec:examples}
We now present example applications of the software package and the formalism. As stated before, we focus on the decay amplitudes \decayamps and its associated filter functions and assume that the unitary errors generated by the frequency shifts \freqshifts are either small (as is the case for gate fidelities) or calibrated out. All of the examples shown below are part of the software documentation as either interactive \jupyter notebooks \cite{Kluyver2016} or Python scripts. In the following, we give angular frequencies and energies in units of inverse times (\eg \si{\per\second}) while ordinary frequencies are given in \si{\hertz} and we write $\ev*{\liouvUe(\tau)} = \liouvUe$ for legibility.

\subsection{Singlet-triplet two-qubit gates}\label{sec:examples:optimized_gates}
In order to benchmark fidelity predictions of our implementation as well as demonstrate its application to nontrivial pulses, we compute the first-order infidelity of the two-qubit gates presented in \citer{Cerfontaine2019} and compare the results to the reference's Monte Carlo calculations. There, a numerically optimized gate set consisting of $\lbrace\mr{X}_{\flatfrac{\pi}{2}}\otimes\mr{I},\mr{Y}_{\flatfrac{\pi}{2}}\otimes\mr{I},\mr{CNOT}\rbrace$ for exchange-coupled singlet-triplet spin qubits is introduced, taking into account different noise spectra and realistic control hardware.

For readers unfamiliar with the reference we briefly summarize the physical system and noise model entering the optimization. The authors consider four electrons confined in a linear array of four quantum dots in a semiconductor heterostructure. Each electron $i\in\lbrace 1,2,3,4\rbrace$ experiences a different static magnetic field $B_i$ so that there is a gradient $b_{ij} = B_i - B_j$ between two adjacent dots $i$ and $j$. This gives rise to spin quantization along the magnetic field axis and defines the eigenstates $\lbrace\ketud,\ketdu\rbrace$ that span the computational subspace of a single qubit so that the accessible Hilbert space of the two-qubit system is spanned by $\lbrace\ketud,\ketdu\rbrace^{\otimes 2}$. The magnitude of the exchange interaction $J_{ij}$ between two adjacent dots $i$ and $j$ is controlled via gate electrodes located on top of the heterostructure that can be pulsed on a nanosecond timescale with an arbitrary waveform generator (AWG). Changing the gate voltages changes the detuning $\eps_{ij}$ of the electrochemical potential between dots and in turn leads to a change in exchange coupling according to the phenomenological model $J_{ij}(\epsilon_{ij})\propto\exp(\epsilon_{ij})$.

The pulses are defined by a set of discrete detuning voltages $\epsilon_{ij}$ passed to an AWG with a sample rate of \SI{1}{\giga S\per\second} and constant magnetic field gradients $b_{ij}$ are assumed. To reflect the fact that the qubits experience a different pulse than what is programmed into the AWG due to cable dispersion and non-ideal control hardware, the detunings are convoluted with an experimental impulse response \cite{Cerfontaine2019}. Finally, the signal is discretized as piecewise constant by slicing each segment into five steps, yielding a time increment of $\Delta t = \SI{0.2}{\nano\second}$.

To find optimal detuning pulses, a Levenberg-Marquardt algorithm iteratively minimizes the infidelity, leakage, and trace distance from the target unitary. For the infidelity, contributions from quasistatic magnetic field noise as well as quasistatic and white charge noise are taken into account during each iteration. Because treating colored (correlated) noise using Monte Carlo methods is computationally expensive (\cf \cref{sec:performance:complexity}), the infidelity due to fast \oneoverf-like noise is only computed for the final gate and not used during the optimization.

Two-qubit interactions are mediated via the exchange $J_{23}$ that makes the states \ketuudd and \ketdduu accessible. They constitute levels outside of the computational subspace that ideally should only be occupied during an entangling gate operation. A non-vanishing population of these states after the operation has ended is therefore unwanted and considered leakage, the magnitude of which we could quantify following \cref{sec:theory:derived_quantities:leakage}. However, here we limit ourselves to determine the infidelity contribution from fast, \viz non-quasistatic, charge noise entering the system through $\epsilon_{ij}$. That is, we consider noise sources $\alpha\in\lbrace\epsilon_{12},\epsilon_{23},\epsilon_{34}\rbrace$. We take the non-linear dependence of the Hamiltonian on the detunings $\epsilon_{ij}$ into account by setting $s_{\epsilon_{ij}}(t) = \pdv*{J_{ij}(\epsilon_{ij}(t))}{\epsilon_{ij}(t)}\propto J_{ij}(\epsilon_{ij}(t))$.

\begin{figure*}[tbp]
    \centering
    \includegraphics[width=\textwidth]{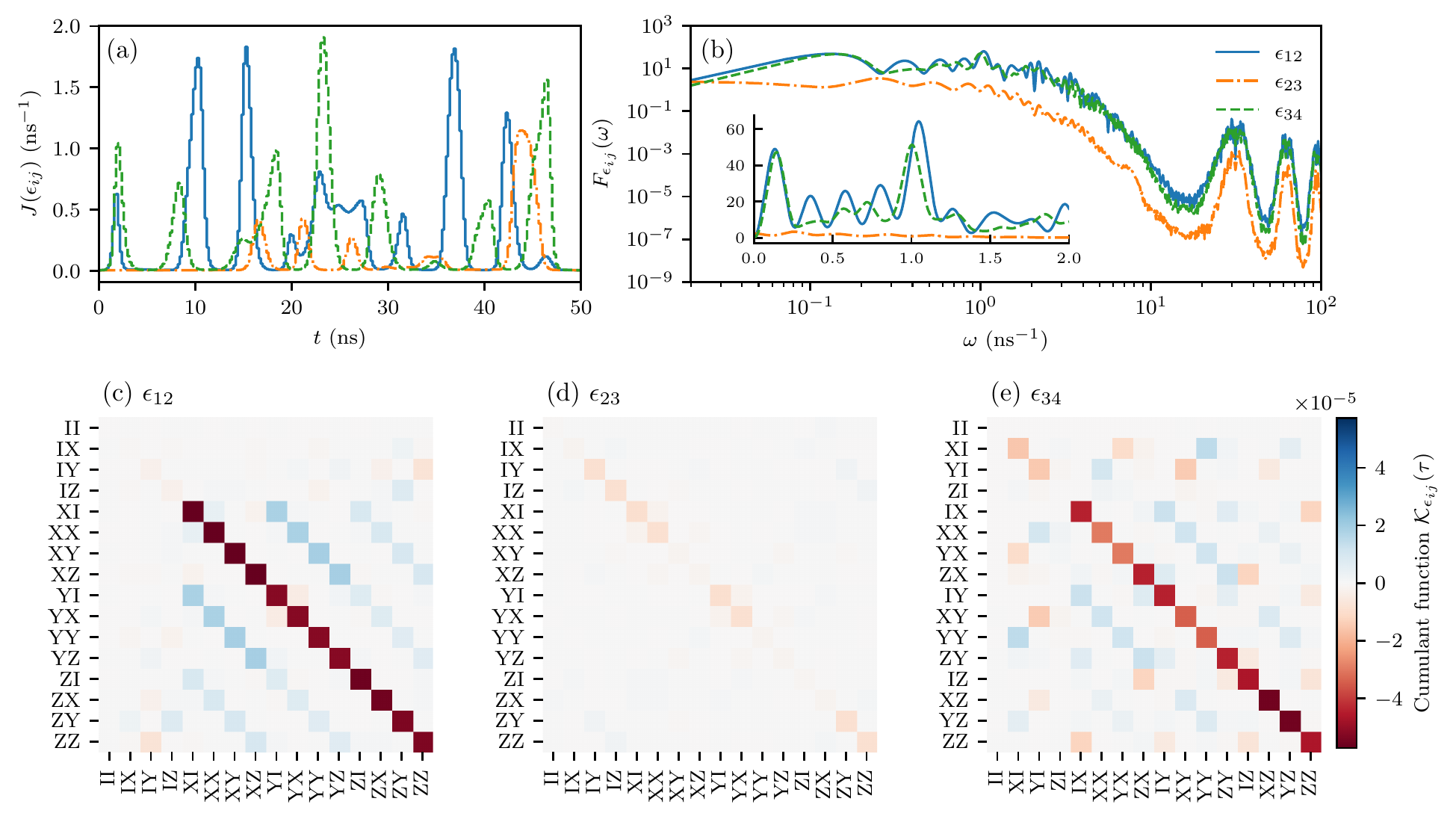}
    \caption{(a) Exchange interaction $J(\epsilon_{ij})$ for the CNOT gate presented in \citer{Cerfontaine2019} as function of time. (b) Filter functions $F_{\epsilon_{ij}}$ for noise in the detunings evaluated on the computational subspace. The filter functions are modulated by oscillations at high frequencies due to numerical artifacts of the finite step size for the time evolution. The inset shows the filter functions in the DC regime on a linear scale with distinct peaks around $\omega = \flatfrac{2\pi}{\tau}$ and $\omega = \flatfrac{50}{\tau}$ ($\tau = \SI{50}{\nano\second}$). (c)--(e) Computational subspace block of the first order approximation of the error transfer matrix, given by the cumulant function $\cumulantfun_{\alpha\alpha}$ excluding second order contributions, for the CNOT gate and the three detunings $\alpha\in\lbrace\epsilon_{12},\epsilon_{23},\epsilon_{34}\rbrace$. Note that in panel (e) the order of the rows and columns was permuted for better comparability.}
    \label{fig:CNOT}
\end{figure*}

\Cref{fig:CNOT} shows the filtered (convoluted) exchange interaction $J_{ij}$ between each pair of dots during the pulse sequence in panel (a) and filter functions plotted as function of frequency in panel (b) for the three different detunings. For a detailed description on how the filter functions were computed in the presence of additional leakage levels refer to \cref{appsec:singlet-triplet}. As one would expect from the fact that the intermediate (inter-qubit) exchange interaction $J_{23}$ (orange dash-dotted lines) is only turned on for short times to entangle the qubits, the filter function for $\eps_{23}$ is smaller by roughly an order of magnitude than the intra-qubit exchange filter functions. Notably, the filter functions for $\eps_{12}$ and $\eps_{34}$ show clear characteristics of DCGs, that is they drop to zero as $\omega\rightarrow 0$, and decouple from quasistatic noise with an error suppression $\propto\omega^2$. This is not unexpected as the optimization minimizes quasistatic noise contributions to the infidelity. In addition, one can also observe small oscillations with period $\SI{5}{\per\nano\second}$ in frequency space that arise as a numerical artifact of the piecewise constant discretization of the control parameters as investigations have shown. If high-frequency spectral components are expected to play a significant role, one needs to be aware of these effects and adjust the simulation parameters appropriately.

The inset of \cref{fig:CNOT}(b) shows the same filter functions for the DC tail on a linear scale. Most notably, $F_{\epsilon_{12}}$ and $F_{\epsilon_{34}}$ have maxima around $\omega = \flatfrac{2\pi}{\tau}$, \ie exactly the frequency matching the pulse duration, and around $\omega = \flatfrac{50}{\tau} = \SI{1}{\per\nano\second}$ with $\tau_\mr{CNOT} = \SI{50}{\nano\second}$. The former is the typical window in which a pulse is most susceptible to noise whereas the latter matches the absolute value of the magnetic field gradients, $b_{12} = -b_{34} = \SI{1}{\per\nano\second}$, indicating that the peak corresponds to the qubit dynamics generated by the magnetic field gradients. Panels (c)--(e) show the cumulant functions $\cumulantfun_{\eps_{ij}}(\tau)$ of the detuning error channels $\eps_{ij}$ on the computational subspace. $\cumulantfun_{\eps_{12}}$ displays clear characteristics of a Pauli channel with only elements on the diagonal and secondary diagonals deviating from zero significantly whereas $\cumulantfun_{\eps_{34}}$ (the target qubit) possesses a more complicated structure.

We now compute the infidelity contribution originating from fast charge noise using \cref{eq:fidelity:avg} but tracing only over the computational subspace to compare to the Monte Carlo calculations of \citer{Cerfontaine2019} (see \cref{appsec:singlet-triplet} for further details). Like the reference, we use a noise spectrum $S_{\epsilon,a}(f)\propto \flatfrac{1}{f^a}$ with $S_{\epsilon,a}(\SI{1}{\MHz}) = \SI{4e-20}{\volt\squared\per\Hz}$ and consider white noise ($a = 0$) and correlated noise with $a = 0.7$ \cite{Dial2013} with infrared and ultraviolet cutoffs $\flatfrac{1}{\tau}$ and $\SI{100}{\per\nano\second}$, respectively. \Cref{tab:infidelities} compares the results in this work with the reference. The values computed here are consistent with the more elaborate Monte Carlo calculations within a few percent. Notably, the deviation is smaller for the smaller noise levels with $a = \num{0.7}$, in line with the fact that we have only computed the contributions from the decay amplitudes \decayamps and thus the leading order perturbation. If we had additionally evaluated the frequency shifts \freqshifts we could have obtained the exact fidelity in the case of Gaussian noise.

\begin{table}
    \centering
    \renewcommand\arraystretch{1.25}
    \newlength{\colwidth}
    \setlength{\colwidth}{1.65 cm}
    \begin{tabular*}{\columnwidth}{l *{4}{S[table-number-alignment=center,table-text-alignment=left,table-format=+1.1e+3,round-mode=figures,round-precision=2,table-column-width=\colwidth]}}
                                                    & \multicolumn{2}{c}{This work}                     & \multicolumn{2}{c}{\citer{Cerfontaine2019}}   \\
    \toprule
        $a$                                         & 0             & \sisetup{round-precision=1} 0.7   & 0             & \sisetup{round-precision=1} 0.7   \\
    \colrule
        $\mr{X}_{\flatfrac{\pi}{2}}\otimes\mr{I}$   & 1.679e-03     & 5.837e-05                         & 1.892e-03     & 5.737e-05                         \\
        $\mr{Y}_{\flatfrac{\pi}{2}}\otimes\mr{I}$   & 1.595e-03     & 5.690e-05                         & 1.689e-03     & 5.622e-05                         \\
        CNOT                                        & 1.498e-03     & 6.399e-05                         & 1.560e-03     & 6.313e-05                         \\
    \botrule
    \end{tabular*}
    \caption{Fast charge noise infidelity contributions to the total average gate fidelity of the two-qubit gate set from \citer{Cerfontaine2019} without capacitive coupling for GaAs \sts qubits compared to the original results. The fidelities are consistent with results from the reference within the uncertainty bounds of \SI{3}{\percent} of the Monte Carlo calculation. The infidelities presented here are all average gate infidelities (cf. \cref{eq:fidelity:avg}, \citerr{Horodecki1999}{Nielsen2002}).}
    \label{tab:infidelities}
\end{table}

\subsection{Rabi driving}\label{sec:examples:rabi_driving}
A widely used method for qubit control is Rabi driving \cite{Wallraff2004,Barends2014,Soare2014,Veldhorst2014}. If we restrict ourselves to the resonant case for simplicity, the control Hamiltonian takes on the general form $\Hc = \flatfrac{\omega_0\pz}{2} + A\sin(\omega_0 t + \phi)\px$. Here, $\omega_0$ is the resonance frequency, $A$ the drive amplitude corresponding to the Rabi frequency in the weak driving limit $\flatfrac{A}{\omega_\mr{0}}\ll 1$, $\Omega_\mr{R}\approx A$, and $\phi$ an adjustable phase giving control over the rotation axis in the $xy$-plane of the Bloch sphere. This Hamiltonian and associated decoherence mechanisms are well-studied in the weak driving regime, where the rotating wave approximation (RWA) can be applied to remove fast-oscillating terms in the rotating frame \cite{Jaynes1963,Gerry2005}. There is a comprehensive understanding of how spectral densities transform to this frame and which frequencies are most relevant to loss of coherence \cite{Yan2013}.

By contrast, the description of a system in the strong driving regime, where $\flatfrac{A}{\omega_0}\sim 1$, is more complicated since the RWA cannot be applied without making large errors. Yet, an improved understanding is desirable because strong driving allows for much shorter gate times and thus shifts the window of relevant noise frequencies towards higher energies where the total noise power is typically lower, \eg for \oneoverf noise. Conversely, faster control also requires more accurate timing to prevent rotation errors. It is therefore of interest to have available tools that can provide a comprehensive picture for Rabi pulses over a wide range of driving amplitudes. By making use of the concatenation property of the filter functions, our formalism can do just that.

The problem that arises when trying to numerically investigate Rabi pulses in the weak driving regime in the lab frame is that typical control operations have a duration $\tau\gg T$ with $T = \flatfrac{2\pi}{\omega_0}$. Since the sampling time step $\Delta t$ should additionally be chosen much smaller than a single drive period in order to sample the time evolution accurately ($\Delta t\ll T$), brute-force simulations are costly.

For $\flatfrac{T}{\Delta t} = 100$ samples per period and assuming Rabi and drive frequencies in typical regimes for SiGe and MOS quantum dots \cite{Zajac2018,Pla2012} or trapped ions \cite{Soare2014}, $\Omega_\mr{R} = \SI{1}{\per\micro\second}$ and $\omega_0 = \SI{20}{\per\nano\second}$, a Monte Carlo simulation of a $\pi$-rotation with approximately \SI{3}{\percent} relative error would require $10^9$ samples in total. Using the filter function formalism, we can drastically reduce the simulation time even beyond the improvement gained from concatenating precomputed filter functions of individual drive periods using \cref{eq:control_matrix:sequence:freq}. This can be achieved with \cref{eq:control_matrix:sequence:periodic:simplified}, which simplifies the calculation of the control matrix for periodic Hamiltonians.

To benchmark our implementation, we use the parameters from above and calculate the control matrix of a NOT gate generated by a Rabi Hamiltonian with three different methods on an \fastprocessor. First, we use \cref{eq:control_matrix:sequence:freq} in a brute force approach. Second, we utilize the concatenation property following \cref{eq:control_matrix:pulse:freq:calculation}. Third, we employ the simplified expression given by \cref{eq:control_matrix:sequence:periodic:simplified}. The brute force approach takes \SI{250}{\second} of wall time whereas calculating the filter function using the standard concatenation is faster by two orders of magnitude, taking \SI{1.5}{\second} to run. Lastly, the calculation utilizing the optimized method is faster again by two orders of magnitude and is completed in \SI{0.056}{\second}.

As an example application, we calculate the filter functions for continuous Rabi driving in the weak and strong driving regimes. For weak driving, we use the parameters from the benchmark above for a pulse of duration $\tau_\mr{weak}\approx\SI{20}{\micro\second}$ that corresponds to \num{20} identity rotations in total. For the strong driving regime, we use the approximate analytical solution for a flux qubit biased at its symmetry point from \citer{Deng2015} with $A = \flatfrac{\omega_0}{4}$ to drive the qubit for $\tau_\mr{strong}\approx\SI{4}{\nano\second}$ so that we achieve the same amount of identity rotations as in the weak driving case. In the reference, strong driving in this regime is shown to give rise to non-negligible counterrotating terms that modulate the Rabi oscillations and which are well-described by Floquet theory applied to the Rabi driving Hamiltonian. While for the regime studied here only two additional modes appear, the results extend to the regime where $A > \omega_0$ and up to eight different frequency components were observed.

\begin{figure}[tbp]
    \centering
    \includegraphics{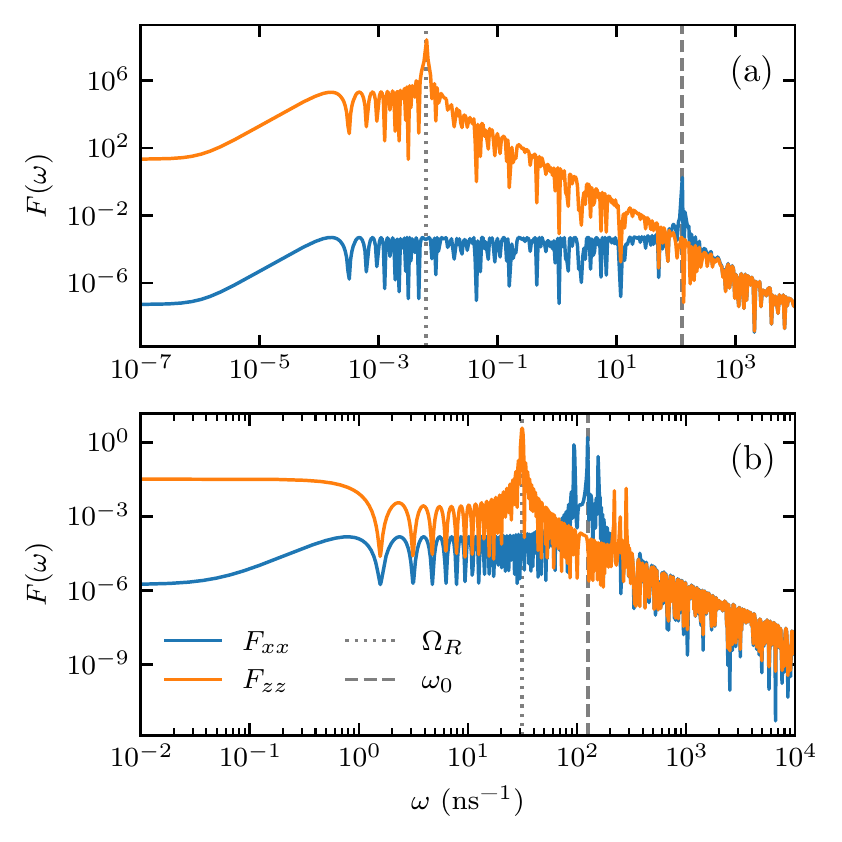}
    \caption{Filter functions for weak (a) and strong (b) Rabi driving (\num{20} identity gates in total). Grey dashed (dotted) lines indicate the respective drive (Rabi) frequencies $\omega_0$ ($\Omega_\mr{R}$). (a) Weak driving with $\flatfrac{A}{\omega_\mr{0}}\ll 1$. The filter function $F_{xx}$ for noise operator \px is approximately constant up to the resonance frequency where it peaks sharply and then aligns with the filter function $F_{zz}$ for \pz. $F_{zz}$ peaks at the Rabi frequency before rolling off with $\omega^{-2}$ and a DC level that is almost ten orders of magnitude larger than the DC level of the transverse filter function $F_{xx}$. (b) Strong driving with $\flatfrac{A}{\omega_\mr{0}}\sim 1$. Again $F_{zz}$ peaks at $\Omega_\mr{R}$ whereas $F_{xx}$ has three distinct peaks at $\omega_\mr{0}$ and $\omega_\mr{0}\pm\Omega_\mr{R}$. These features also appear at slightly higher frequencies in $F_{zz}$ due to the strong coupling.}
    \label{fig:filter_function:rabi:weak_vs_strong}
\end{figure}

\Cref{fig:filter_function:rabi:weak_vs_strong} shows the filter functions $F_{xx}$ and $F_{zz}$ for the \px and \pz noise operators in the weak (a) and the strong (b) driving regime. Both display sharp peaks at their Rabi frequencies and the resonance frequency for $F_{zz}$ and $F_{xx}$, respectively. We expect these features as they correspond to perturbations of the qubit Hamiltonian that are resonant with the qubit dynamics about an axis orthogonal to them. For weak driving, $F_{xx}$ is constant up to the resonance frequency where it peaks sharply and then aligns with $F_{zz}$. The latter has a peak at the Rabi frequency before rolling off with $\omega^{-2}$ and a DC level that is almost ten orders of magnitude larger than that of the transverse filter function. This behavior is consistent with the results by \citet{Yan2013}, who show that the noise sources dominating decoherence during driven evolution are $S_{xx}(\omega_0)$ and $S_{zz}(\Omega_\mr{R})$. Note that the piecewise constant control approximation causes the weak driving filter functions to level off towards low frequencies after an initial roll-off (here at $\omega\sim\SI{1}{\per\milli\second}$). By decreasing the discretization time step $\Delta t$, one can shift the frequency at which this effect occurs to lower frequencies and thus attribute the feature to a numerical artefact of the approximation. However, the decoupling properties depend quite sensitively on the pulse duration.

In case of strong driving, the two filter functions are closer in amplitude for lower frequencies. In addition, $F_{xx}$ also peaks at $\omega = \omega_0\pm\Omega_\mr{R}$. These peaks also show up at higher frequencies in the dephasing filter function $F_{zz}$, reflecting frequency mixing in the strong coupling regime.
While both filter functions show characteristics of a DCG in the weak driving regime, that is they drop to zero as $\omega\rightarrow 0$, this is not the case in the strong driving regime. Instead, there they approach a constant level for small frequencies. On top of rotation errors from timing inaccuracies, we may thus expect naive strong driving gates to be more susceptible to quasistatic noise than weak driving gates. By shaping the pulse envelope of the strong driving gate the decoupling properties could be recovered.

\subsection{Randomized Benchmarking}\label{sec:examples:randomized_benchmarking}
Standard Randomized Benchmarking (SRB) and related methods, for example interleaved RB, are popular tools to assess the quality of a qubit system and the operations used to control it \cite{Knill2008,Magesan2011,Magesan2012}. The basic protocol consists of constructing $K$ random sequences of varying length $m$ of gates drawn from the Clifford group  
\footnote{The Clifford group is a subgroup of the special unitary group with the advantage that compositions are easy to compute and that averaging over all unitaries can under reasonable assumptions be replaced by averaging over all Cliffords. This makes the Clifford gates a convenient choice for benchmarking. For a nice, short introduction as well as further references, see \cite{Ozols2008}},
and appending a final inversion gate so that the identity operation should be performed in total. Each of these pulse sequences is applied to an initial state $\kpsi$ in order to measure the survival probability $p(\kpsi)$ after the sequence. In reality, the applied operations are subject to noise and experimental imprecisions. This renders them imperfect and results in a survival probability smaller than one. Assuming gate-independent errors, the average gate fidelity $\avgfid$ is then obtained by fitting the measured survival probabilities for each sequence length to the zeroth-order exponential model \cite{Magesan2011}
\begin{equation}\label{eq:SRB}
    p(\kpsi) = A \left(1 - \frac{dr}{d-1}\right)^m + B,
\end{equation}
where $r = 1 - \avgfid$ is the average error per single gate to be extracted from the fit, $A$ and $B$ are parameters capturing state preparation and measurement (SPAM) errors, and $d$ is the dimensionality of the system.

One of the main assumptions of the SRB protocol is that temporal correlations of the noise are small on timescales longer than the average gate time \cite{Magesan2011}. If this requirement is not satisfied, \eg if \oneoverf noise plays a dominant role, the decay of the sequence fidelity can have non-exponential components \cite{Epstein2014,Fogarty2015,Feng2016} and a single exponential fit will not produce the true average gate fidelity \cite{Mavadia2018,Edmunds2020}. The filter function formalism suggests itself to numerically probe RB experiments in such systems for two reasons. First, it enables the study of gate performance subject to noise with correlation times longer than individual gate times. This regime, where a simple description in terms of individual, isolated quantum operations fails, is accessible in the filter function formalism because universal classical noise can be included by the power spectral density $S(\omega)$. Second, the simulation of a RB experiment can be performed efficiently by using the concatenation property. Because RB sequences are compiled from a limited set of gates whose filter functions may be precomputed, one only needs to concatenate $m$ filter functions for a single sequence of length $m$ to gain access to the survival probability.

Since for sufficiently long RB sequences $r\in\order{1}$, and we would need to include the frequency shifts \freqshifts in a full simulation following \cref{eq:cumulant_expansion} because the low-noise approximation \cref{eq:error_transfer_matrix:approx} does not hold in this regime. Unfortunately, the concatenation property does not hold for \freqshifts. Therefore, we focus on the high-fidelity regime where the exponential decay of the sequence fidelity may be approximated to linear order and only the decay amplitudes \decayamps need to be considered.

In order to evaluate the survival probability of a RB experiment using filter functions, we employ the state fidelity from \cref{sec:theory:derived_quantities:state_fidelity-measurements} and focus on the single-qubit case with $d = 2$ and the (normalized) Pauli basis from \cref{eq:basis:pauli}. Because the ideal action of a RB sequence is the identity we have $\liouvQ = \eye$. Assuming we prepare and measure in the computational basis, $\kpsi\in\lbrace\ket{0}, \ket{1}\rbrace$ so that $\sqrt{2}\dket{\rho} = \dket{\sigma_0}\pm\dket{\sigma_3}$, we simplify \cref{eq:fidelity:state} to
\begin{equation}\label{eq:fidelity:state:RB}
    \begin{split}
        \fid(\kpsi, \liouvU_\mr{RB}(\op{\psi})) &= \frac{1}{2}\bigl(\liouvUe_{00} +
                                                                \liouvUe_{33}\pm
                                                                \liouvUe_{03}\pm
                                                                \liouvUe_{30}\bigr) \\
                                                &= \frac{1 + \liouvUe_{33}}{2} \approx 1 - \frac{1}{2}\sum_{k\neq 3}\decayamps_{kk}.
    \end{split}
\end{equation}
For the second equality we used that \liouvUe is trace-preserving and unital (\cf \cref{sec:theory:transfer_matrix:derivation}) while in the last step we approximated the expression using \cref{eq:error_transfer_matrix:approx,eq:cumulant:truncated:liouville:pauli}. For our simulation, we neglect SPAM errors so that $A =  B =  0.5$, choose $\kpsi = \ket{0}$, and approximate \cref{eq:SRB} as
\begin{equation}\label{eq:fidelity:state:RB:fit}
    p(\ket{0}) = \fid(\ket{0}, \liouvU_\mr{RB}(\op{0}))\approx 1 - rm
\end{equation}
for small gate errors $r\ll 1$ since this is the regime which we can efficiently simulate using the concatenation property.

We simulate single-qubit SRB experiments using three different gate sets to generate the 24 elements of the Clifford group. For the first gate set we implement the group by naive \enquote{single} rotations about the symmetry axes of the cube. Each pulse corresponds to a single time segment during which one rotation is performed so that the $j$-th element is given by $Q_j = \exp(-\i\phi_j\vec{n}_j\cdot\vec{\sigma})$. We compile the other two gate sets from primitive $\flatfrac{\pi}{2}$ $x$- and $y$-rotations so that on average each Clifford gate consists of \num{3.75} primitive gates (see \citer{Cerfontaine2019ex}). For the specific implementation of the primitive $\flatfrac{\pi}{2}$-gates we compare \enquote{naive} rotations, \ie with a single time segment so that $Q_j = \exp(\flatfrac{-\i\pi\sigma_j}{4})$ for $j\in\lbrace x, y\rbrace$, and the \enquote{optimized} gates from \citer{Cerfontaine2019}. Pulse durations are chosen such that the average duration of all 24 Clifford gates generated from a single gate set is equal for all three gate sets. This is to ensure that the different implementations of the Clifford gates are sensitive to the same noise frequencies.

We investigate white noise and correlated noise with $S(\omega)\propto\omega^{-0.7}$ assuming the same noise spectrum on each Cartesian axis of the Bloch sphere and normalize the noise power for each gate set and noise type (white and correlated) so that the average Clifford infidelity $r$ is the same throughout. We then randomly draw $K = \num{100}$ sequences for \num{11} different lengths $m\in[1, 101]$ and concatenate the $m$ Clifford gates using \cref{eq:control_matrix:sequence:freq} to compute the control matrix of the entire sequence. For the integral in \cref{eq:decay_amplitudes:freq} we choose the ultraviolet cutoff frequency two orders of magnitude above the inverse duration of the shortest pulse, $f_\mr{UV} = \flatfrac{10^2}{\tau_\mr{min}}$. Similarly, the infrared cutoff is chosen as $f_\mr{IR} = \flatfrac{10^{-2}}{m_\mr{max}\tau_\mr{max}}$ with $m_\mr{max} = 101$ and $\tau_\mr{max} = 7\tau_\mr{min}$ (since the longest gate is compiled from seven primitive gates with duration $\tau_\mr{min}$) to guarantee that all nontrivial structure of the filter functions is resolved at small frequencies
\footnote{For a precise fidelity estimate, the infrared cutoff should be extended to $f=0$. However, we are only interested in a qualitative picture and neglect this part of the spectrum here. At frequencies much smaller than $\approx\flatfrac{1}{\tau}$ where $\tau$ is the duration of the entire control operation, the filter function is constant and we therefore do not disregard any interesting features by setting $f_\mr{IR} = \flatfrac{10^{-2}}{\tau} = \flatfrac{10^{-2}}{m_\mr{max}\tau_\mr{max}}$.}. Finally, we fit \cref{eq:fidelity:state:RB:fit} to the infidelities computed for the different noise spectra.

\begin{figure}[tbp]
    \centering
    \includegraphics{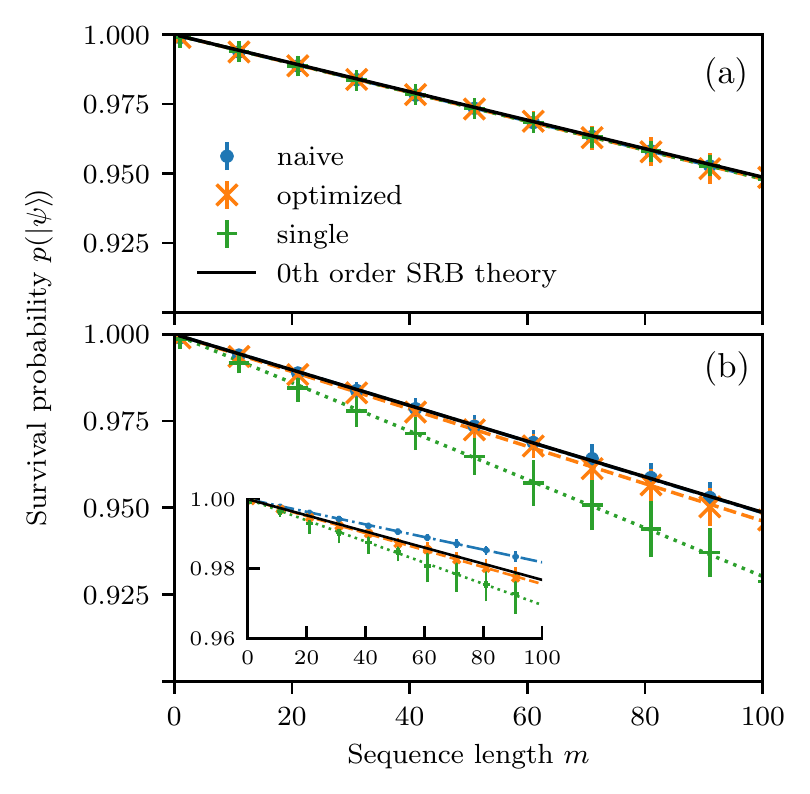}
    \caption{Simulation of a Standard \gls{rb} (SRB) experiment using \num{100} random sequences per point for different gate and noise types (see the main text for an explanation of the gate type monikers). Dashed lines are fits of \cref{eq:fidelity:state:RB:fit} to the data while the solid black lines correspond to a zeroth-order SRB model with $A=B=\num{0.5}$ and the true average gate infidelity per Clifford $r$. Errorbars show the standard deviation of the SRB sequence fidelities, illustrating that for the \enquote{single} gate set noise correlations can lead to amplified destructive and constructive interference of errors. The same noise spectrum is used for all three error channels (\px, \py, \pz) and the large plots show the sum of all contributions. (a) Uncorrelated white noise with the noise power adjusted for each gate type so that the average error per gate $r$ is constant over all gate types. No notable deviation is seen between different gate types. (b) Correlated \oneoverf-like noise with noise power adjusted to match the average Clifford fidelity in (a). The decay of the \enquote{single} gateset differs considerably from that of the other gate sets and the SRB decay expected for the given average gate fidelity, whereas \enquote{naive} and \enquote{optimized} gates match the zeroth order SRB model well, indicating that correlations in the noise affect the relation between SRB decay and average gate fidelity in a gateset-dependent way. Inset: contributions from \pz-noise show that the sequence fidelity can be better than expected for certain gate types and noise channels.}
    \label{fig:randomized_benchmarking:noise_comparison}
\end{figure}

The results of the simulation are shown in \cref{fig:randomized_benchmarking:noise_comparison} (a) and (b) for white and correlated noise, respectively. For white noise, the survival probability agrees well with the SRB prediction for all gate types whereas for \oneoverf-like noise the \enquote{single} gates (green pluses) deviate considerably. Hence, fitting the zeroth-order SRB model to such data will not reveal the true average gate fidelity although errors are of order unity. We note that \citerr{Epstein2014}{Ball2016} found similar results using different methods for \oneoverf and perfectly correlated DC noise, respectively. The former observed SRB to estimate $r$ within \SI{25}{\percent} and the latter found the mean of the SRB fidelity distribution to deviate from the mode, thereby giving rise to incorrectly estimated fidelities.

On top of affirming the findings by the references, our results demonstrate that the accuracy of the predictions made by SRB theory, \ie that the \gls{rb} decay rate directly corresponds to the average error rate of the gates, not only depends on the gate implementation but also on which error channels are assumed. This can be seen from the inset of \cref{fig:randomized_benchmarking:noise_comparison}(b), where only dephasing noise (\pz) contributions are shown. For this noise channel and the \enquote{naive} gates, one finds a slower \gls{rb} decay than expected from the actual average gate fidelity, so that the latter would be overestimated by an \gls{rb} experiment, whereas the \enquote{single} gates show the opposite behavior. Depending on the gate set and relevant error channels, non-Markovian noise may thus even lead to improved sequence fidelities due to errors interfering destructively. This behavior is captured by the pulse correlation filter functions whose contributions to the sequence fidelity lead to the deviations from the SRB prediction.

Notably, the data for the \enquote{optimized} gates agree with the prediction for every noise channel individually which implies that correlations between pulses are suppressed. This highlights the formalism's attractiveness for numerical gate optimization as the pulse correlation filter functions $F\gth{gg'}(\omega)$ may be exploited to suppress correlation errors. To be more explicit, the correlation decay amplitudes $\decayamps\gth{gg'}$ from \cref{eq:decay_amplitudes:pulse_correlation} can be used to construct cost functions for quantum optimal control algorithms like GRAPE \cite{Khaneja2005,Schulte-Herbruggen2005} or CRAB \cite{Caneva2011}. By constructing linear combinations of $\decayamps\gth{gg'}$ with different pulse indices $g$ and $g'$, correlations between any number of pulses can be specifically targeted and suppressed using numerical pulse optimization.

\subsection{Quantum Fourier transform}\label{sec:examples:qft}
To demonstrate the flexibility of our software implementation, we calculate filter functions for a four-qubit \acrfull{qft} \cite{Coppersmith1994,NielsenChuang2011} circuit. QFT plays an important role in many quantum algorithms such as Shor's algorithm \cite{Shor1997} and quantum phase estimation \cite{NielsenChuang2011}. For the underlying gate set, we assume a standard Rabi driving model with IQ control and nearest neighbor exchange. That is, we assume full control of the $x$- and $y$-axes of the individual qubits as well as the exchange interaction mediating coupling between two neighboring qubits. This system allows for native access to the minimal gateset $\mathbb{G} = \lbrace\mr{X}_{i}(\flatfrac{\pi}{2}),\mr{Y}_{i}(\flatfrac{\pi}{2}),\mr{CR}_{ij}(\flatfrac{\pi}{2^3})\rbrace$ where $\mr{CR}_{ij}(\phi)$ denotes a controlled rotation by $\phi$ about $z$ with control qubit $i$ and target qubit $j$. Controlled-$z$ rotations by angles $\flatfrac{\pi}{2^m}$ as required for the QFT can thus be obtained by concatenating $2^{3-m}$ minimal gates $\mr{CR}_{ij}(\flatfrac{\pi}{2^3})$.

\begin{figure*}[tbp]
    \includegraphics[width=0.965\textwidth]{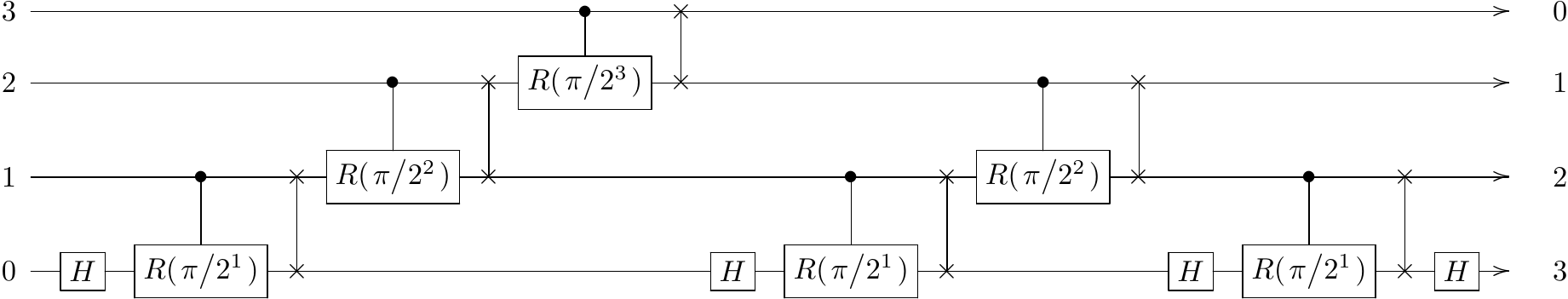}
    \includegraphics[width=\textwidth]{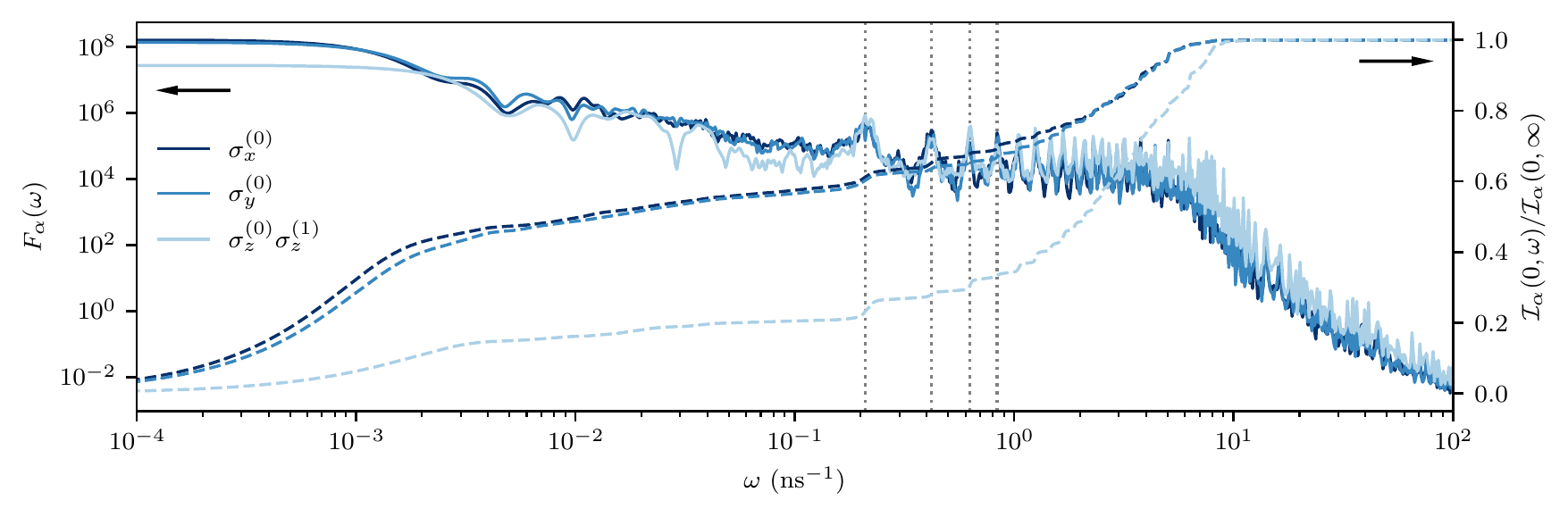}
    \caption{Top: Circuit for a \gls{qft} on four qubits with nearest-neighbor coupling. Labels next to the wires indicate the qubit index, showing that the final SWAP operation has already been carried out. Bottom: Filter functions for noise operators on the first qubit ($i = 0$). Dotted grey lines indicate the positions of the $n$-th harmonic, $\omega_n = \flatfrac{2\pi n}{\tau}$ with $\tau = \SI{30}{\nano\second}$ the duration of the gates in $\mathbb{G}$, for $n\in\lbrace 1, 2, 3, 4\rbrace$. The filter functions have a baseline of around $10^4$ in the range $\omega\in[10^{-1}, 10^{1}]$ \si{\per\nano\second} before they drop down to follow the usual $1/\omega^2$ behavior. The dashed lines show the error sensitivities $\mc{I}_\alpha(\omega_1,\omega_2)\coloneqq\int_{\omega_1}^{\omega_2}\dd{\omega} F_\alpha(\omega)$ in the frequency band $[0, \omega]$ as a fraction of the total sensitivity $\mc{I}_\alpha(0,\infty)$. These are closely related to the entanglement fidelity (\cf \cref{eq:filter_function:fidelity,eq:infidelity:ent:integral}) and suggest that high frequencies up to the knee at $\omega\approx\SI{10}{\per\nano\second}$ cannot be neglected if the cutoff frequency of the noise is sufficiently high or the spectrum does not drop off quickly enough (note the linear scale as opposed to the logarithmic scale for the filter functions).}
    \label{fig:qft}
\end{figure*}

Despite native access to all necessary gates, we employ \qutip's implementation \cite{Johansson2013} of the GRAPE algorithm \cite{Khaneja2005,Schulte-Herbruggen2005} to generate the gates in order to highlight our method's suitability for numerically optimized pulses. For the optimization we choose a time step of $\Delta t = \SI{1}{\nano\second}$ and a total gate duration of $\tau = \SI{30}{\nano\second}$. For completeness, see \cref{appsec:qft} for details on the optimized gates. We then construct the remaining required gates by sequencing these elementary gates, \ie the Hadamard gate $\mr{H}_i = \mr{X}_{i}(\flatfrac{\pi}{2})\circ\mr{X}_{i}(\flatfrac{\pi}{2})\circ\mr{Y}_{i}(\flatfrac{\pi}{2})$, where $\mr{B}\circ\mr{A}$ denotes the composition of gates A and B such that gate A is executed before gate B. To map the canonical circuit \cite{NielsenChuang2011} onto our specific qubit layout with only nearest-neighbor coupling, we furthermore introduce SWAP operations to couple distant qubits. These gates can be implemented by three CNOTs, $\mr{SWAP}_{ij} = \mr{CNOT}_{ij}\circ\mr{CNOT}_{ji}\circ\mr{CNOT}_{ij}$. The CNOTs in turn are obtained by a Hadamard transform of the controlled phase gate, $\mr{CNOT}_{ij} = \mr{H}_j\circ\mr{CR}_{ij}(\pi)\circ\mr{H}_j$. The complete quantum circuit is shown at the top of \cref{fig:qft}; for the canonical circuit with all-to-all connectivity refer to \citer{NielsenChuang2011}. In total, there are \num{442} elementary pulses, \num{198} of which are required for the three SWAPs on the first two qubits, so that the entire algorithm would take $\sim\SI{13}{\micro\second}$ to run. Note that the circuit could be compressed in time by parallelizing some operations but for simplicity we only execute gates sequentially and do not execute dedicated idling gates.

In order to leverage the extensibility of the filter function approach (see \cref{sec:performance:extending_hilbert_spaces}), we use a Pauli basis for the pulses and proceed as follows:
\begin{enumerate}
    \item Instantiate the \pulsesequence objects for the elementary gates $\mathbb{G}$ for the first two qubits and cache the control matrices.
    \item Compile all required single- and two-qubit pulses by concatenating the \pulsesequences that implement $\mathbb{G}$.
    \item Extend the \pulsesequences to the full four-qubit Hilbert space.
    \item Recursively concatenate recurring gate sequences by concatenating four-qubit \pulsesequences, \eg $\mr{SWAP}_{10}\circ\mr{CR_{10}}(\flatfrac{\pi}{2^1})\circ\mr{H}_0$, in order to optimally use the performance benefit offered by \cref{eq:control_matrix:sequence:freq}
    \item Concatenate the last \pulsesequences to get the complete QFT pulse.
\end{enumerate}
For our gate parameters and \num{400} frequency points, this procedure takes around \SI{5}{\second} on an \fastprocessor, whereas computing the filter functions naively using \cref{eq:control_matrix:pulse:freq:calculation} takes around \SI{4}{\minute}. The resulting filter functions are shown in \cref{fig:qft} for the noise operators affecting the first qubit; for an in-depth discussion and validation of the fidelities predicted, see the accompanying letter \citer{Cerfontaine2021} and its supplementary information. Evidently, the fidelity of the algorithm is most susceptible to DC noise; below roughly $\omega\lessapprox 10^{-3}\,\si{\per\nano\second}$ the filter functions level off at their maximum value. In the \si{\giga\hertz} range there is a plateau with sharp peaks corresponding to the $n$-th harmonics of the inverse pulse duration $\omega_n = \flatfrac{2\pi n}{\tau}$, where the leftmost belongs to $n=1$. The dashed lines show the error sensitivities $\mc{I}_\alpha(\omega_1, \omega_2)\coloneqq\int_{\omega_1}^{\omega_2}\dd{\omega}F(\omega)$ in the frequency band $[0, \omega]$ relative to the total sensitivity $\mc{I}_\alpha(0,\infty)$. For a white spectrum, \ie $S(\omega)=\mr{const.}$, this quantifies the fraction of the total entanglement infidelity that is accumulated up to frequency $\omega$ (\cf \cref{eq:filter_function:fidelity,eq:infidelity:ent:integral}). Thus, to obtain a precise estimate of the algorithm's fidelity, five frequency decades need to be taken into account.

These insights demonstrate that our method represents a useful tool to analyze how and to which degree small algorithms are affected by correlated errors, and how this effect depends on the gate implementation. It could thus also be used to choose or optimize gates in an algorithm-specific way.

\section{Further Considerations}\label{sec:considerata}
Before we conclude, let us address two possible avenues for future work, one for the formalism itself and one for its application.

To extend our approach to the filter function formalism beyond the scope discussed in this work, the most evident path forward is to allow for quantum mechanical baths instead of purely classical ones. Such an extension would facilitate studying for example non-unital $T_1$-like processes. In fact, the filter function formalism was originally introduced considering quantum baths such as spin-boson models \cite{Martinis2003,Uhrig2007} or more general baths \cite{Kofman2001,Yuge2011,Paz-Silva2017}, but it remains an open question whether this can be applied to our presentation of the formalism and the numerical implementation in particular. In a fully quantum-mechanical treatment, (sufficiently weak) noise coupling into the quantum system can be modelled via a set of bath operators $\{D_\alpha(t)\}_\alpha$ so that $\Hn(t) = \sum_\alpha\Ba(t)\otimes D_\alpha(t)$ (the classical case is recovered by replacing $D_\alpha(t)\rightarrow b_\alpha(t)\eye$) \cite{Breuer2007}. Accordingly, the ensemble average over the stochastic bath variables $\{b_\alpha(t)\}_\alpha$ needs to be replaced by the quantum expectation value $\tr_B(\placeholder\rho_B)$ with respect to the state $\rho_B$ of the bath $B$. One therefore needs to deal with correlation functions of bath operators instead of stochastic variables. An immediate consequence for numerical applications is hence an increased dimensionality of the system, which could be dealt with by using analytical expressions for the partial trace over the bath.

For future applications of our method, it would be interesting to study the effects of noise correlations in quantum error correction (QEC) schemes \cite{Devitt2013,Ng2011,Nickerson2019}. While extensive research has been performed on QEC, noise is usually assumed to be uncorrelated between error correction cycles. In this respect, our formalism may shed light on effects that need to be taken into account for a realistic description of the protocol. As outlined above, we can compute expectation values of (stabilizer) measurements in a straightforward manner from the error transfer matrix. Unfortunately, this implies performing the ensemble average over different noise realizations, therefore removing all correlations between subsequent measurement outcomes for a given noise realization. Hence, the same feature that allows us to calculate the quantum process for correlated noise, namely that we compute only the final map by averaging over all \enquote{paths} leading to it, prevents us from studying correlations between consecutive cycles. To overcome this limitation in the context of quantum memory one could invoke the principle of deferred measurement \cite{NielsenChuang2011} and move all measurements to the end of the circuit, replacing classically controlled operations dependent on the measurement outcomes by conditional quantum operations. Alternatively, to incorporate the probabilistic nature of measurements, one could devise a branching model that implements the classically controlled recovery operation by following both conditional branches of measurement outcomes with weights corresponding to the measurement probabilities as computed from the ensemble-averaged error transfer matrix. An intriguing connection also exists to the quantum Zeno effect, for which quantum systems subject to periodic projective measurements have been identified with a filter function \cite{Kofman2000,Kofman2001,Chaudhry2016}.

\section{Conclusion and Outlook}\label{sec:conclusion}
As quantum control schemes become more sophisticated and take into account realistic hardware constraints and sequencing effects, their analytic description becomes cumbersome, making numerical tools invaluable for analyzing pulse performance. In the above, we have shown that the filter function formalism lends itself naturally to these tasks since the central objects of our formulation, the interaction picture noise operators, obey a simple composition rule which can be utilized to efficiently calculate them for a sequence of quantum gates. Because the nature of the noise is encoded in a power spectral density in the frequency domain, its effects are isolated from the description of the control until they are evaluated by the overlap integral of noise spectrum and filter function. Hence, the noise operators are highly reusable in calculations and can serve as an economic way of simulating pulse sequences.

Building on the results of a separate publication \cite{Cerfontaine2021}, we have presented a general framework to study decoherence mechanisms and pulse correlations in quantum systems coupled to generic classical noise environments. By combining the quantum operations and filter function formalisms, we have shown how to compute the Liouville representation of the exact error channel of an arbitrary control operation in the presence of Gaussian noise. For non-Gaussian noise our results become perturbative in the noise strength. Furthermore, we have introduced the \filterfunctions \python software package that implements the aforementioned method. We showed both analytically and numerically that our software implementation can outperform Monte Carlo techniques by orders of magnitude. By employing the formalism and software to study several examples we demonstrated the wide range of possible applications.

The capacity for applications in quantum optimal control has already been established above. In a forthcoming publication, we will present analytical derivatives for the fidelity filter function, \cref{eq:filter_function:fidelity}, and their implementation in the software package \cite{Le2021}. Together with the infidelity, \cref{eq:infidelity:ent:integral}, they can serve as efficient cost functions for pulse optimization in the presence of realistic, correlated noise \cite{qopt,Teske2021}. Since our method offers insight into correlations between pulses at different positions in a sequence, the pulse correlation filter function $F\gth{gg'}(\omega)$ with $g\neq g'$ can additionally serve as a tool for studying under which conditions pulses decouple from noise with long correlation times. Such insight would be valuable to design pulses for algorithms. Another interesting application could be quantum error correction in the regime of long-time correlated noise as outlined above in \cref{sec:considerata}, where we also briefly touched upon a possible extension of the framework to quantum mechanical baths.

The tools presented here, both analytical and numerical as implemented in the \filterfunctions software package \cite{software}, provide an accessible way for computing filter functions in generic control settings across the different material platforms employed in quantum technologies and beyond.

\begin{acknowledgments}
This work was supported by the European Research Council (ERC) under the European Union's Horizon 2020 research and innovation program (Grant Agreement No. 679342).
\end{acknowledgments}

\onecolumngrid
\appendix
\section{Additional derivations}\label{appsec:derivations:cumulant:pauli}
In this appendix we show additional derivations omitted from the main text.
\subsection{Derivation of the single-qubit cumulant function in the Liouville representation}
For a single qubit, the Pauli basis $\left\lbrace\sigma_i\right\rbrace_{i=0}^3 = \flatfrac{\left\lbrace\eye,\px,\py,\pz\right\rbrace}{\sqrt{2}}$ is a natural choice to define the Liouville representation. In this case, the trace tensor \cref{eq:trace_tensor} can be simplified and thus \cref{eq:cumulant:truncated:liouville} given a more intuitive form which we derive in this appendix. Since the cumulant function is linear in the noise indices $\alpha,\beta$ we drop them in the following for legibility. Our results hold for both a single pair of noise indices and the total cumulant. We start by observing the relation
\begin{equation}\label{eq:trace_tensor:four_paulis}
    T_{klij} = \tr(\sigma_k\sigma_l\sigma_i\sigma_j) = \flatfrac{(\delta_{kl}\delta_{ij} - \delta_{ki}\delta_{lj} + \delta_{kj}\delta_{ji})}{2}
\end{equation}
for the Pauli basis elements $\sigma_k, k\in\lbrace 1, 2, 3\rbrace$. Including the identity element $\sigma_0$ in the trace tensor gives additional terms. However, as we show now none of these contribute to \cumulantfun because they cancel out.

First, since the noise Hamiltonian $\Hn(t)$ is traceless and therefore $\ctrlmat_{\alpha 0}(t) = 0$, we have $\decayamps_{kl},\freqshifts_{kl}\propto (1 - \delta_{k0})(1 - \delta_{0l})$, \ie the first column and row of both the decay amplitude and frequency shift matrices are zero, and hence terms in the sum of \cref{eq:cumulant:truncated:liouville} with either $k = 0$ or $l = 0$ vanish. Next, for $i = j = 0$ all of the traces cancel out as can be easily seen. The last possible cases are given by $i = 0, j\neq 0$ and vice versa. For these cases we have
\begin{equation}\label{eq:trace_tensor:three_paulis}
    T_{kl0j} = T_{klj0} = \frac{1}{\sqrt{2}}\tr(\sigma_k\sigma_l\sigma_j) = \frac{\i}{2}\varepsilon_{klj}
\end{equation}
with $\varepsilon_{kli}$ the completely antisymmetric tensor. Both of the above cases vanish in \cumulantfun since, taking the case $j = 0$ for example,
\begin{subequations}\label{eq:trace_tensor:three_paulis:aggregate}
\begin{equation} \label{eq:trace_tensor:three_paulis:aggregate:decayamps}
    \frac{1}{2}\left(T_{kl0i} - T_{k0li} - T_{kil0} + T_{ki0l}\right) =
    \frac{\i}{2}\left(\varepsilon_{klj} - \varepsilon_{lkj} - \varepsilon_{klj} + \varepsilon_{lkj}\right) = 0
\end{equation}
for the decay amplitudes \decayamps and
\begin{equation} \label{eq:trace_tensor:three_paulis:aggregate:freqshifts}
    \frac{1}{2}\left(T_{kl0i} - T_{lk0i} - T_{kli0} + T_{lki0}\right) =
    \frac{\i}{2}\left(\varepsilon_{kli} - \varepsilon_{lki} - \varepsilon_{kli} + \varepsilon_{lki}\right) = 0
\end{equation}
\end{subequations}
for the frequency shifts \freqshifts. Hence, only terms with $i,j > 0$ contribute and we can plug the simplified expressions for the trace tensor $T_{klij}$, \cref{eq:trace_tensor:four_paulis}, into \cref{eq:cumulant:truncated:liouville} to write the cumulant function for a single qubit and the Pauli basis concisely as
\begin{align}
    \cumulantfun_{ij}(\tau) &= -\frac{1}{2}\sum_{kl}\biggl(\freqshifts_{kl}\left(T_{klji} - T_{lkji} - T_{klij} + T_{lkij}\right)
                                               + \decayamps_{kl}\left(T_{klji} - T_{kjli} - T_{kilj} + T_{kijl}\right)\biggr) \\
                            &= -\sum_{kl}\biggl(\freqshifts_{kl}(\delta_{ki}\delta_{lj} - \delta_{kj}\delta_{li})
                                               + \decayamps_{kl}(\delta_{kl}\delta_{ij} - \delta_{kj}\delta_{li})\biggr) \\
                            &= \freqshifts_{ji} - \freqshifts_{ij} + \decayamps_{ij} - \delta_{ij}\tr\decayamps \\
                            &= \begin{cases}
                                  - \sum_{k\neq i}\decayamps_{kk}                           &\qif* i = j,   \\
                                  - \freqshifts_{ij} + \freqshifts_{ji} + \decayamps_{ij}   &\qif* i\neq j,
                               \end{cases}
\end{align}
as given in the main text.

\subsection{Evaluation of the integrals in \cref{eq:frequency_shifts:freq}}\label{appsec:derivations:frequency_shifts:integral}
Here we calculate the integrals appearing in the calculation of the frequency shifts \freqshifts, \cref{eq:frequency_shifts:integral}, given by
\begin{equation}
    I_{ijmn}\gth{g}(\omega) = \int_{t_{g-1}}^{t_g}\dd{t}\e^{\i\Omega_{ij}\gth{g}(t - t_{g-1}) - \i\omega t}
                              \int_{t_{g-1}}^{t}\dd{t'}\e^{\i\Omega_{mn}\gth{g}(t' - t_{g-1}) + \i\omega t'}.
\end{equation}
The inner integration is simple to perform and we get
\begin{equation}
    I_{ijmn}\gth{g}(\omega) = \int_{t_{g-1}}^{t_g}\dd{t}\e^{\i\Omega_{ij}\gth{g}(t - t_{g-1}) - \i\omega(t - t_{g-1})}\times
    \begin{cases}
        \frac{\e^{\i(\omega + \Omega_{mn}\gth{g})(t - t_{g-1})} - 1}{\i(\omega + \Omega_{mn}\gth{g})}   &\qif* \omega + \Omega_{mn}\gth{g}\neq 0 \\
        t - t_{g-1}                                                                                     &\qif* \omega + \Omega_{mn}\gth{g} = 0.
    \end{cases}
\end{equation}
Shifting the limits of integration and performing integration by parts in the case $\omega + \Omega_{mn}\gth{g} = 0$ then yields
\begin{equation}
    I_{ijmn}\gth{g}(\omega) = \begin{cases}
        \frac{1}{\omega + \Omega_{mn}\gth{g}}\left(
            \frac{\e^{\i(\Omega_{ij}\gth{g} - \omega)\Delta t_g} - 1}{\Omega_{ij}\gth{g} - \omega} -
            \frac{\e^{\i(\Omega_{ij}\gth{g} + \Omega_{mn}\gth{g})\Delta t_g} - 1}{\Omega_{ij}\gth{g} + \Omega_{mn}\gth{g}}
        \right) &\qif* \omega + \Omega_{mn}\gth{g}\neq 0, \\
        \frac{1}{\Omega_{ij}\gth{g} - \omega}\left(
            \frac{\e^{\i(\Omega_{ij}\gth{g} - \omega)\Delta t_g} - 1}{\Omega_{ij}\gth{g} - \omega} -
            \i\Delta t_g\e^{\i(\Omega_{ij}\gth{g} - \omega)\Delta t_g}
        \right) &\qif* \omega + \Omega_{mn}\gth{g} = 0 \wedge \Omega_{ij}\gth{g} - \omega\neq 0, \\
        \flatfrac{\Delta t_g^2}{2} &\qif* \omega + \Omega_{mn}\gth{g} = 0 \wedge \Omega_{ij}\gth{g} - \omega = 0.
    \end{cases}
\end{equation}

\subsection{Simplifying the calculation of the entanglement infidelity}\label{appsec:derivations:fidelity}
In the main text, we claimed that the contribution of noise sources $(\alpha,\beta)$ to the total entanglement infidelity $\entinfid(\liouvUe) = \sum_{\alpha\beta}\infid_{\alpha\beta}$ reduces from the trace of the cumulant function \cumulantfun to
\begin{align}
    \infid_{\alpha\beta} &= -\frac{1}{d^2}\tr\cumulantfun_{\alpha\beta} \label{appeq:infid} \\
                         &= \frac{1}{d}\tr\decayamps_{\alpha\beta}.
\end{align}
To show this, we substitute \cumulantfun by its definition in terms of \freqshifts and \decayamps according to \cref{eq:cumulant:truncated:liouville}. This yields for the trace
\begin{equation}\label{appeq:cumulant:trace:1}
    \begin{split}
        \tr\cumulantfun_{\alpha\beta} &= -\frac{1}{2}\sum_{kl}\delta_{ij}(f_{ijkl}\freqshifts_{\alpha\beta,kl} + g_{ijkl}\decayamps_{\alpha\beta,kl}) \\
                                      &= -\frac{1}{2}\sum_{ikl}\decayamps_{\alpha\beta,kl}\left(T_{klii} + T_{lkii} - 2 T_{kili}\right)
    \end{split}
\end{equation}
since \freqshifts is antisymmetric. In order to further simplify the trace tensors on the right hand side of \cref{appeq:cumulant:trace:1}, we observe that the orthonormality and completeness of the operator basis \basis defining the Liouville representation of \cumulantfun (\cf \cref{eq:basis}) is equivalent to requiring that $\basis\ad\basis = \eye$ with \basis reshaped into a $d^2\times d^2$ matrix by a suitable mapping. This condition may also be written as
\begin{equation}\label{eq:basis:identity}
\begin{split}
    \delta_{ac}\delta_{bd} &= \sum_{k} C^\ast_{k,ab} C_{k,cd} \\
                           &= \sum_{k} C_{k,ba} C_{k,cd}
\end{split}
\end{equation}
because every element $C_k$ is Hermitian. Using this relation in \cref{appeq:cumulant:trace:1} then yields
\begin{equation}\label{appeq:cumulant:trace:2}
    \begin{split}
        \tr\cumulantfun_{\alpha\beta} &= -\frac{1}{2}\sum_{kl}\decayamps_{\alpha\beta,kl}\left(2d\delta_{kl} - 2\tr(C_k)\tr(C_l)\right) \\
                                      &= -d\tr\decayamps_{\alpha\beta}.
    \end{split}
\end{equation}
The last equality only holds true for bases with a single non-traceless element (the identity), such as the bases discussed in \cref{sec:performance:basis}. This is because in this case, $\tr(C_k) = 0$ for $k > 0$ whereas $\decayamps_{\alpha\beta,kl} = 0$ for either $k = 0$ or $l = 0$ since \decayamps is a function of the traceless noise Hamiltonian for which $\tr(C_0\Hn) \propto \tr\Hn = 0$ (\ie the first column of the control matrix is zero, see \cref{eq:control_matrix,eq:decay_amplitudes:time}). Finally, substituting \cref{appeq:cumulant:trace:2} into \cref{appeq:infid} we obtain our result
\begin{equation}
    \infid_{\alpha\beta} = \frac{1}{d}\tr\decayamps_{\alpha\beta}.
\end{equation}
\section{Singlet-Triplet Gate Fidelity}\label{appsec:singlet-triplet}
In this appendix we lay out in more detail how the fidelity of the optimized \sts qubit gates from \citer{Cerfontaine2019} was calculated using filter functions. In two singlet-triplet qubits, angular momentum conservation suppresses occupancy of states with non-vanishing magnetic spin quantum number $m_s$ so that the total accessible state space of dimension $d=6$ is spanned by $\lbrace\ketudud,\ketuddu,\ketduud,\ketdudu,\ketuudd,\ketdduu\rbrace$. A straightforward method to single out the computational subspace (CS) dynamics from those on the whole space would be to simply project the error transfer matrix $\liouvUe\approx\eye + \cumulantfun$ with \cumulantfun the cumulant function onto the CS as proposed by \citet{Wood2018}, that is calculate the fidelity as $\entfid = \mr{tr}\bigl(\Pi_c\liouvUe\bigr)/d_c^2$ where $\Pi_c$ is the Liouville representation of the projector onto the CS and $d_c = 4$ the dimension of the CS. However, here we use a more involved procedure in order to gain more insight from the error transfer matrix as well as to obtain a better comparison to the fidelities computed by \citet{Cerfontaine2019}, who map the final $6\times 6$ propagator to the closest unitary on the $4\times 4$ CS during their Monte Carlo simulation.

To calculate the fidelity of the target unitary on the $4\times 4$ CS, we thus construct an orthonormal operator basis \basis of the full $6\times 6$ space that is partitioned into elements which are nontrivial only on the CS on the one hand and elements which are nontrivial only on the remaining space on the other such that $\basis = \basis^c\cup\basis^\ell$. Using such a basis, we can then trace only over CS elements of the error transfer matrix \liouvUe in \cref{eq:fidelity:ent} to obtain the fidelity of the gate on the CS. Moreover, we retain the opportunity to characterize the gates on the basis of the Pauli matrices.

Since there is no obvious way to extend the Pauli basis for two qubits to the complete space we proceed as follows: For the CS, we pad the two-qubit Pauli basis with zeros on the leakage levels, \ie
\begin{equation}\label{eq:basis:cnot}
    C_i^c\doteq\bordermatrix{~                       &     & \footnotesize{\ketuudd} & \footnotesize{\ketdduu} \cr
                                                     & P_i & 0                       & 0                       \cr
                             \footnotesize{\brauudd} & 0   & 0                       & 0                       \cr
                             \footnotesize{\bradduu} & 0   & 0                       & 0                       \cr}
    \qcomma{i\in\{0,\dotsc,15\}},
\end{equation}
where the $P_i$ are normalized two-qubit Pauli matrices (\cf \cref{eq:basis:pauli}) in the basis $\lbrace\ketudud,\ketuddu,\ketduud,\ketdudu\rbrace$. To complete the basis we require an additional 20 elements orthogonal to the 16 padded Pauli matrices. We obtain the remaining elements by first expanding the $C_i^c$ in an arbitrary basis $\left\lbrace\Lambda_i\right\rbrace_{i=0}^{35}$ of the complete space (we choose a GGM, \cf \cref{eq:basis:ggm}, for simplicity), yielding a $16\times 36$ matrix of expansion coefficients:
\begin{equation}
    M_{ij} = \tr(C_i^c\Lambda_j).
\end{equation}
We then compute an orthonormal vector basis $V$ (a matrix of size $36\times 20$) for the null space of $M$ using singular value decomposition $M = U\Sigma V\ad$ and acquire the corresponding basis matrices as
\begin{equation}
    C_i^\ell = \sum_j\Lambda_j V_{ji}\qcomma{i\in\lbrace 0,\dotsc,19\rbrace}.
\end{equation}
Finally, to account for the fact that \citer{Cerfontaine2019} map the total propagator to the closest unitary on the CS, we exclude the identity Pauli element $C_0^c\propto\text{diag}(1, 1, 1, 1, 0, 0)$ from the trace over the computational subspace part of \liouvUe represented in the basis $\basis = \basis^c\cup\basis^\ell$ when calculating the fidelity,
\begin{equation}
    \entfid = \frac{1}{16}\sum_{i=1}^{15}\liouvUe_{ii},
\end{equation}
since for unitary operations on the CS we have $\cumulantfun_{00} \approx 1 - \liouvUe_{00} = 1 - \mr{tr}\bigl(C_0^c\Ue C_0^c\Ue\ad\bigr) = 0$. Hence, excluding $\liouvUe_{00}$ from the trace corresponds to partially disregarding non-unitary components of the error channel on the computational subspace. Although not the only element that differs compared to the closest subspace unitary, $\cumulantfun_{00}$ contains the most obvious contribution, whereas those of other elements are more difficult to disentangle into unitary and non-unitary components.

\begin{figure}
    \centering
    \includegraphics{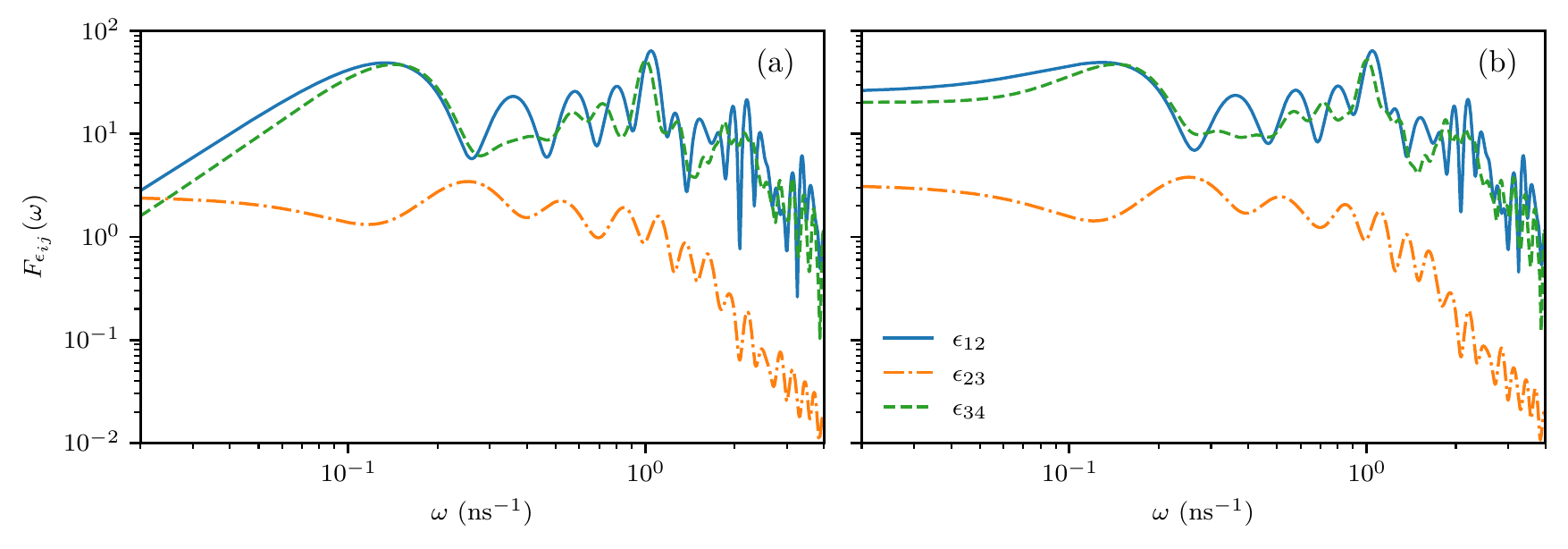}
    \caption{Filter functions of the voltage detunings $\epsilon_{ij}$ excluding (a) and including (b) the zero-padded identity matrix basis element $C_0^c\propto\text{diag}(1,1,1,1,0,0)$ for the computational subspace. Evidently, including $C_0^c$ removes the DCG character, namely that $F_{\epsilon_{ij}}(\omega)\rightarrow 0$ as $\omega\rightarrow 0$, of the gates but has little effect on the high-frequency behavior. As the pulse optimization minimizes, among other figures of merit, the infidelity of the final propagator mapped to the closest unitary on the computational subspace due to quasistatic and fast white noise, this indicates that excluding $C_0^c$ from the filter function corresponds to partially neglecting non-unitary components of the propagator on the computational subspace.}
    \label{appfig:filter_function:cnot}
\end{figure}

Similar to the fidelity, we also obtain the canonical filter function shown in panel (b) of \cref{fig:CNOT} by summing only over columns one through 15 of the control matrix, $F_{\epsilon_{ij}}(\omega) = \sum_{k=1}^{15}\bigl\lvert\ctrlmat_{\epsilon_{ij} k}(\omega)\bigr\rvert^2$. In fact, including the first column, corresponding to the padded identity matrix $C_0^c$, in the filter function removes the DCG character of $F_{\epsilon_{12}}(\omega)$ and $F_{\epsilon_{34}}(\omega)$, which instead approach a constant level of around 20 (note that the filter function is dimensionless in our units) at zero frequency. This is consistent with the fact that the gates were optimized using quasistatic and fast white noise contributions to the fidelity after mapping to the closest unitary on the computational subspace. We have performed Monte Carlo resimulations that support this reading. In \cref{appfig:filter_function:cnot} we show the filter functions once including and once excluding the contributions from $C_0^c$.

\section{GRAPE-optimized gate set for QFT}\label{appsec:qft}
\begin{figure}
    \centering
    \includegraphics{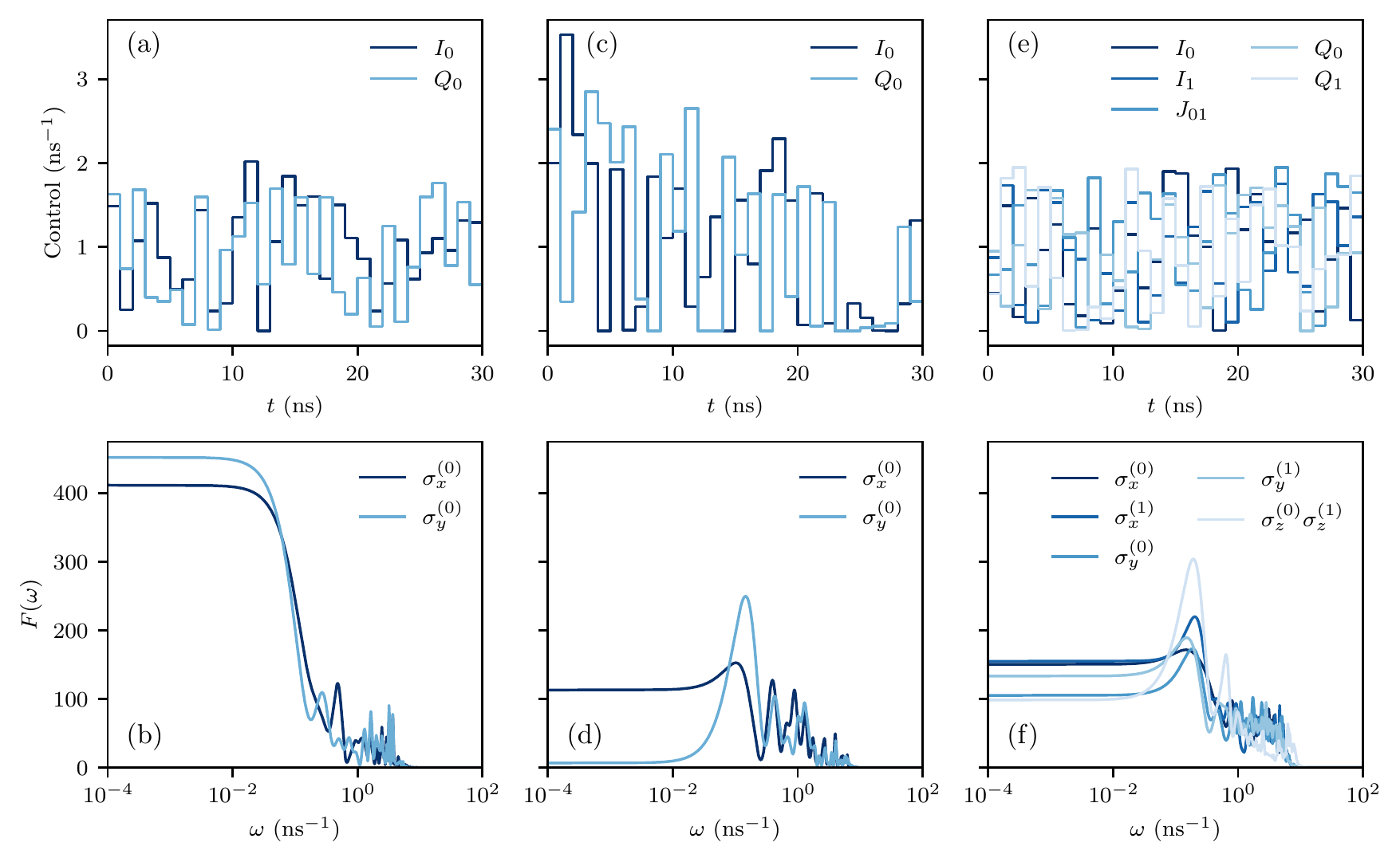}
    \caption{Control fields (top row) and corresponding filter functions (bottom row) of the GRAPE-optimized pulses in $\mathbb{G}$. (a),(b) $\text{X}_0(\flatfrac{\pi}{2})$; (c),(d) $\text{Y}_0(\flatfrac{\pi}{2})$; (e),(f) $\text{CR}_{01}(\flatfrac{\pi}{2^3})$. Note that the optimization is neither very sophisticated nor realistic as the algorithm only maximizes the systematic (coherent) fidelity $\mr{tr}\bigl(UQ\ad_\mr{targ}\bigr)/d$ and the randomly distributed initial control amplitudes are not subject to any constraints.}
    \label{appfig:qft:gates}
\end{figure}

For completeness, in this appendix we give details on the GRAPE-optimized pulses for the gate set $\mathbb{G} = \lbrace\mr{X}_{i}(\flatfrac{\pi}{2}),\mr{Y}_{i}(\flatfrac{\pi}{2}),\mr{CR}_{ij}(\flatfrac{\pi}{2^3})\rbrace$ used in \cref{sec:examples:qft} to simulate a \gls{qft} algorithm. As mentioned in the main text, we consider a toy Rabi driving model with IQ single-qubit control and exchange to mediate inter-qubit coupling. Cast in the language of quantum optimal control theory this translates to a vanishing drift (static) Hamiltonian, $H_\mr{d} =  0$, and a control Hamiltonian in the rotating frame given by
\begin{gather}
    \Hc(t) = \Hc\gth{0}(t)\otimes\eye + \eye\otimes\Hc\gth{1}(t) + \Hc\gth{01}(t), \\
    \Hc\gth{i}(t) = I_i(t)\px\gth{i} + Q_i(t)\py\gth{i},\qquad\Hc\gth{ij}(t) = J_{ij}(t)\pz\gth{i}\otimes\pz\gth{j},
\end{gather}
where $I_i(t)$ and $Q_i(t)$ are the in-phase and quadrature pulse envelopes and $\sigma_{x,y}\gth{i}$ are the Pauli matrices acting on the $i$-th and extended trivially to the other qubit. As our goal is only of illustrative nature and not to provide a detailed gate optimization, we obtain the controls $\lbrace I_0(t), Q_0(t), I_1(t), Q_1(t), J_{12}(t)\rbrace$ for the gate set $\mathbb{G}$ using the GRAPE algorithm implemented in \qutip \cite{Johansson2013} initialized with randomly distributed amplitudes. The resulting pulses and the corresponding filter functions for the relevant noise operators are shown in \cref{appfig:qft:gates}.

\section{Convergence Bounds}\label{appsec:convergence}
In this appendix we give bounds for the convergence of the expansions employed in the main text for the case of purely auto-correlated noise, $S_{\alpha\beta}(\omega) = \delta_{\alpha\beta}S_{\alpha\beta}(\omega) =  S_\alpha(\omega)$, following the approach by \citet{Green2013}. For Gaussian noise, our expansion is exact when including first and second order \acrfull{me} terms. Hence, the convergence radius of the \gls{me} becomes infinite and the fidelity can be computed exactly by evaluating the matrix exponential \cref{eq:cumulant}. For non-Gaussian noise, the following considerations apply.
\subsection{Magnus Expansion}\label{appsec:convergence:magnus_expansion}
The \gls{me} of the error propagator \cref{eq:magnus_expansion:1} converges if $\int_0^\tau\dd{t}\norm*{\Hnt(t)} < \pi$ with $\norm{A}^2 = \dotHS{A}{A} = \sum_{ij}\abs{A_{ij}}^2$ the Frobenius norm \cite{Moan1999}. We assume a time dependence of the noise operators of the form $\Ba(t) = s_\alpha(t)\Ba$. By the Cauchy-Schwarz inequality we then have
\begin{equation}
    \begin{split}
        \bigl\lVert\Hnt(t)\bigr\rVert^2 &= \norm{\Hn(t)}^2 \\
                          &= \sum_{\alpha\beta} s_\alpha(t) s_{\beta}(t) b_\alpha(t) b_{\beta}(t)\dotHS{B_\alpha}{B_{\beta}} \\
                          &\leq\sum_{\alpha\beta} s_\alpha(t) s_{\beta}(t) b_\alpha(t) b_{\beta}(t)
                             \norm{B_\alpha}\norm{B_{\beta}} \\
                          &\leq\biggl[\sum_{\alpha}\sum_{g=1}^{G}\vartheta^{(g)}(t)
                             s_\alpha^{(g)}b_\alpha^{(m)}\norm{B_\alpha}\biggr]^2
    \end{split}
\end{equation}
where $b_{\alpha}^{(m)}$ is the maximum value that the noise assumes during the pulse, $\vartheta\gth{g}(t) = \Theta(t - t_{g-1}) - \Theta(t - t_g)$ is one during the $g$-th time interval and zero else, and where we approximated the time evolution as piecewise constant. Then, in order to guarantee convergence of the \gls{me},
\begin{equation}
    \begin{split}
        \int_0^{\tau}\dd{t}\norm*{\Hnt(t)} &\leq\int_0^\tau\dd{t}\abs\bigg{\sum_{\alpha}\sum_{g=1}^{G}
                                              \vartheta^{(g)}(t) s_\alpha^{(g)} b_\alpha^{(m)}\norm{B_\alpha}} \\
                                           &= \sum_{\alpha} b_\alpha^{(m)}\norm{B_\alpha}\sum_{g=1}^{G}s_\alpha^{(g)}
                                              \int_{t_{g-1}}^{t_{g}}\dd{t} \\
                                           &= \sum_{\alpha} C_m\delta b_\alpha\norm{B_\alpha}\sum_{g=1}^{G}
                                              s_\alpha^{(g)}\Delta t_g \\
                                           &\eqqcolon N
    \end{split}
\end{equation}
where we have expressed the in principle unknown maximum noise amplitude $b_\alpha^{(m)}$ in terms of the root mean square value $\delta b_\alpha$. That is, $b_\alpha^{(m)} = C_m \expval*{b_\alpha(0)^2}^{1/2} =  C_m \delta b_\alpha$ for a sufficiently large value $C_m$. Finally, realizing that $\delta b_\alpha^2 = \int\frac{\dd{\omega}}{2\pi} S_\alpha(\omega)$ and by the triangle inequality,
\begin{equation}
    \begin{split}
        N &= C_m\sum_{\alpha}\norm{B_\alpha}\biggl[\int_{-\infty}^\infty\frac{\mathrm{d}\omega}{2\pi} S_\alpha(\omega)\biggr]^{1/2}
            \sum_{g=1}^{G} s_\alpha^{(g)}\Delta t_g \\
          &\leq C_m\biggl[\sum_{\alpha}\norm{B_\alpha}^2\int_{-\infty}^\infty\frac{\mathrm{d}\omega}{2\pi} S_\alpha(\omega)
            \biggl(\sum_{g=1}^{G} s_\alpha^{(g)}\Delta t_g\biggr)^2\biggr]^{1/2} \\
          &\eqqcolon C_m\xi \\
          &\overset{!}{<} \pi
    \end{split}
\end{equation}
where we have introduced the parameter $\xi$. Thus, the expansion converges if $\xi < \flatfrac{\pi}{C_m}$. However, we note that in practice the rms noise amplitude $\delta b_\alpha$ will often be infinite, limiting the usefulness of this bound for certain noise spectra.
\subsection{Infidelity}\label{appsec:convergence:infidelity}
Again assuming a time dependence $\Ba(t) = s_\alpha(t)\Ba$ as well as piecewise constant control, we note that for the infidelity we have (\cf \cref{eq:fidelity:ent})
\begin{equation}
    \begin{split}
        \abs{\tr(\decayamps)} &= \abs\Bigg{\sum_{\alpha}\int_0^\tau\dd{t_2}\int_0^\tau\dd{t_1}
                                 \expval{b_\alpha(t_1)b_\alpha(t_2)}\sum_{k}\ctrlmat_{\alpha k}(t_1)\ctrlmat_{\alpha k}(t_2)} \\
                              &\leq\abs\Bigg{\sum_{\alpha}\int_0^\tau\dd{t_2}\int_0^\tau\dd{t_1}
                                 \expval{b_\alpha(t_1)b_\alpha(t_2)}\sum_{g,g'=1}^{G}\vartheta^{(g)}(t_1)\vartheta^{(g')}(t_2)
                                 s_\alpha^{(g)} s_\alpha^{(g')} \norm{B_\alpha}^2} \\
                              &\leq\sum_{\alpha}\norm{B_\alpha}^2
                                 \underbrace{\expval{b_\alpha^2(0)}}_{\int\frac{\dd{\omega}}{2\pi}S_\alpha(\omega)}
                                 \sum_{g,g'=1}^{G} s_\alpha^{(g)} s_\alpha^{(g')}
                                 \abs\Bigg{\int_{t_{g'-1}}^{t_{g'}}\dd{t_2}\int_{t_{g-1}}^{t_g}\dd{t_1}
                                 \underbrace{\overline{\expval{b_\alpha(t_1)b_\alpha(t_2)}}}_{\abs{\placeholder}\leq 1}} \\
                              &\leq\sum_{\alpha}\left[\norm{B_\alpha}^2
                                 \int_{-\infty}^\infty\frac{\mathrm{d}\omega}{2\pi}S_\alpha(\omega)
                                 \biggl(\sum_{g=1}^{G}s_\alpha^{(g)}\Delta t_g\biggr)^2\right] \\
                              &= \xi^2,
    \end{split}
\end{equation}
where, going from the second to the third line, we have factored out the total power of noise source $\alpha$ from the cross-correlation function, $\expval{b_\alpha(t_1) b_\alpha(t_2)} = \expval{b_\alpha^2(0)}\bigl\lvert\overline{\expval{b_\alpha(t_1)b_\alpha(t_2)}}\bigr\rvert$. Thus, the first order infidelity \cref{eq:fidelity:ent} is upper bounded by $\flatfrac{\xi^2}{d}$, the same parameter also bounding the convergence of the \gls{me}, and higher orders can be neglected if $\xi^2\ll 1$.

Note that similar arguments can be made for the higher orders of the \gls{me} \cite{Green2013}. In particular, the $n$-th order \gls{me} term containing $n$-point correlation functions of the noise is of order $\order{\xi^n}$ as stated in the main text.

\twocolumngrid

\end{document}